\documentclass[letterpaper,10pt,twocolumn]{IEEEtran}
\makeatletter
\def\ps@headings{%
\def\@oddhead{\mbox{}\scriptsize\rightmark \hfil \thepage}%
\def\@evenhead{\scriptsize\thepage \hfil \leftmark\mbox{}}%
\def\@oddfoot{}%
\def\@evenfoot{}}
\makeatother
\usepackage{amssymb,amsmath,amsfonts,bm,epsfig,graphicx,theorem}
\usepackage{rotating,setspace,latexsym,epsf,color,epstopdf}
\usepackage{cite,subfigure, algorithm, algorithmic}
\usepackage{dsfont}

\newcommand{\Xv}{\mathbf{X}}
\newcommand{\xv}{\mathbf{x}}
\newcommand{\Yv}{\mathbf{Y}}

\newcommand{\yv}{\mathbf{y}}

\newcommand{\Qv}{\mathbf{Q}}
\newcommand{\Tv}{\mathbf{T}}
\newcommand{\Pd}{\mathbb{P}}
\newcommand{\Pb}{\mathbb{P}}

\newcommand{\Vc}{\mathcal{V}}
\newcommand{\Ec}{\mathcal{E}}
\newcommand{\Nc}{\mathcal{N}}
\newcommand{\Xc}{\mathcal{X}}
\newcommand{\Ac}{\mathcal{A}}

\newcommand{\Cc}{\mathcal{C}}

\newcommand{\Sc}{\mathcal{S}}
\newcommand{\Kc}{\mathcal{K}}
\newcommand{\ov}{\mathbf{0}}
\newcommand{\ev}{\mathbf{e}}
\newcommand{\Tc}{\mathcal{T}}

\newcommand{\tEq}{\!=\!} 

\newtheorem{Theorem}{Theorem}

\newtheorem{Lemma}{Lemma}

\newtheorem{Definition}{Definition}
\newtheorem{Assumption}{Assumption}
\newenvironment{Proof}[1]{\medskip\par\noindent
{\bf Proof:\,}\,#1}{{\mbox{\,$\blacksquare$}\par}}

\begin{document}


\title{Learning the Interference Graph of a Wireless Network}
\author{Jing~Yang,~\IEEEmembership{Member,~IEEE,}
  ~Stark~C.~Draper,~\IEEEmembership{Senior Member,~IEEE,}
  ~Robert~Nowak,~\IEEEmembership{Fellow,~IEEE}

  \thanks{This work was supported, in part, by the Air Force Office of
    Scientific Research under grant FA9550-09-1-0140, by the National
    Science Foundation under grants CCF-0963834, ECCS-1405403, ECCS-1650299, and by the National
    Sciences and Engineering Research Council of Canada through a
    Discovery Grant.  This material was presented, in part, at the
    {\em IEEE Int. Symp. Inf. Theory}, July 2012
    \cite{yang_isit12}.}   
    \thanks{J. Yang is with the School of
    Electrical Engineering and Computer Science, The Pennsylvania
    State University, University Park, PA, 16802, USA. Email:
    yangjing@psu.edu.}  \thanks{S. C. Draper is with the Department of
    Electrical and Computer Engineering, University of Toronto,
    Canada. Email: stark.draper@utoronto.ca. } \thanks{R. Nowak is
    with the Department of Electrical and Computer Engineering,
    University of Wisconsin-Madison, WI, 53706, USA. Email:
    nowak@ece.wisc.edu}}

\maketitle

\begin{abstract}
A key challenge in wireless networking is the management of
interference between transmissions. Identifying which transmitters
interfere with each other is a crucial first step.  In this paper we
cast the task of estimating the a wireless interference environment as
a graph learning problem. Nodes represent transmitters and edges
represent the presence of interference between pairs of transmitters.
We passively observe network traffic transmission patterns and collect
information on transmission successes and failures.  We establish
bounds on the number of observations (each a snapshot of a network
traffic pattern) required to identify the interference graph reliably
with high probability.

Our main results are scaling laws that tell us how the number of
observations must grow in terms of the total number of nodes $n$ in
the network and the maximum number of interfering transmitters $d$ per
node (maximum node degree). The effects of hidden terminal
interference (i.e., interference not detectable via carrier sensing)
on the observation requirements are also quantified. We show that to
identify the graph it is necessary and sufficient that the observation
period grows like $d^2 \log n$, and we propose a practical algorithm
that reliably identifies the graph from this length of
observation. The observation requirements scale quite mildly with
network size, and networks with sparse interference (small $d$) can be
identified more rapidly. Computational experiments based on a
realistic simulations of the traffic and protocol lend additional
support to these conclusions.
\end{abstract}



\section{Introduction}\label{sec:intro}
Due to the broadcast nature of wireless communications, simultaneous
transmissions in the same frequency band and time slot may interfere
with each other, thereby limiting system throughput. Interference
estimation is thus an essential part of wireless network
operation. Knowledge of interference among nodes is an important input
in many wireless network configuration tasks, such as channel
assignment, transmit power control, and scheduling.

A number of recent efforts have made significant progress towards the
goal of real-time identification of the network interference
environment. Some of the recent approaches (e.g., Interference maps
\cite{int_map} and Micro-probing \cite{int_probe}) inject traffic into
the network to infer occurrences of interference. While such
approaches can be quite accurate in determining interference, the
overhead of making such active measurements in a large network limits
their practicality. In particular, the periodic use of active probing
methods to identify interference can place a significant burden on the
network when network conditions change over time. Time variations can
be caused by changes in the physical environment, e.g., by an office
door being left open, or by mobility among the clients, or the dynamic
power configuration in heterogeneous networks~\cite{LiuEtAl:13}.

The desire to avoid the overhead of active probing motivates the
development of passive techniques such as the ``Passive Interference
Estimation'' (PIE) algorithm~\cite{vivek_thesis,pie_nsdi}. Inspired by
two passive WLAN (wireless local area network) monitoring approaches
(Jigsaw \cite{jigsaw1, jigsaw2} and WIT \cite{wit}) PIE infers
interference structure from passive observation of the pattern of
successful, and unsuccessful, transmissions. Experimental studies in
small testbeds~\cite{pie_nsdi} show that PIE is quite promising, but
very little is understood about how the method scales up to large
complex networks.  Understanding such scaling is the focus of this
paper.

We formulate passive interference estimation as a statistical learning
problem. Given an arbitrary WLAN that consists of $n$ access points
(APs) and a variable number of mobile clients, our goal is to recover
the ``interference'' or ``conflict'' graph among these APs with as few
measurements as possible.  This graph encodes the interference
relations between APs and other APs' clients.  We study two versions
of the problem.  In the first version we study ``direct'' interference
between APs where edges in the graph indicate that a pair of APs are
within each other's carrier sensing range.  Letting $d$ be the maximum
number of interfering APs per AP, we show that to identify the
conflict graph one must collect a number of measurements proportional
to $d^2 \log n$. This is quite mild dependence on the network size $n$
and indicates that interference graph inference is scalable to large
networks and that sparser patterns of interference are easier to
identify than denser patterns.  In the second version we quantify the
effect of ``hidden'' terminal interference.  This type of interference
occurs when one AP interferes with another AP's clients, but the
transmission of the interfering AP is not detectable by the other AP.
In this case feedback on transmission successes and failures is
required to estimate the graph. For both versions of the problem we
present easy-to-implement graph estimation algorithms. The algorithms
are adaptive to $d$, in the sense that they do not require apriori
knowledge of $d$.  We also develop lower bounds that demonstrate that
the time-complexity attained by the algorithms cannot be improved upon
by any other scheme. This provides insight into the time scale over
which network interference patterns can be identified and tracked in
dynamically changing networks.


The inference problem studied in this paper is somewhat
  reminiscent of network tomography, in which one attempts to identify
  network parameters based on network topology-dependent
  measurements~\cite{Castro04}. In \cite{Duffield:2002,
    Castro:2004:LBH}, knowledge of pairwise metric values, such as
  end-to-end loss, are used to identify the network
  topology. In~\cite{SattariFM13} a network tomography approach based
  on network coding is discussed. The idea is to exploit the
  topology-dependent correlation introduced by network coding in the
  content of received packets to reverse-engineer the topology. For
  the interference graph inference problem studied in this paper, the
  topology information is encoded in the transmission patterns and
  feedback information. We use pairwise relationships of an AP's
  transmission status to recover the direct inference graph, and use
  the feedback information together with the transmission patterns to
  recover the hidden inference graph.

The problem of inferring the hidden interference graph studied in
  this paper is related to group testing problems. Group testing is a
  process to identify a small set of defective items from a large
  population through a sequence of tests.  Each test is conducted on a
  subset of all items, where a positive outcome indicates that at
  least one defective item is contained in the subset. One major
  research direction in group testing is the design of the test
  matrix, and the number of tests required to effect reliable
  detection. In~\cite{AtiaS12}, the group testing problem is
  formulated as a channel coding/decoding problem. The test matrix is
  formed randomly, and the total number of tests required to identify
  the defective set is characterized from an information theoretic
  perspective. In~\cite{Cheraghchi:2012:GGT} group testing on graphs
  is investigated.  Each test must conform to the constraints imposed
  by the graph. A distributed group testing algorithm which detects
  the set of defective sensors from binary messages exchanged by the
  sensors is studied in~\cite{TosIc:2013:DSF}. The inference of a
  hidden inference graph studied in this paper is similar to a group
  testing problem in the sense that, for each AP, its hidden
  interferers form the set of ``distinguished" APs from the whole set
  of APs. If AP $i$ transmits at time $t$, the subset of APs that are
  transmitting simultaneously at time $t$ form the test pool, and the
  feedback information is the test outcome. If the feedback
  information indicates that the transmission of AP $i$ fails, it
  implies that at least one hidden interferer is within the test pool;
  otherwise, no hidden interferer is included in the pool. Identifying
  the hidden interference graph is equivalent to recover the set of
  ``distinguished" APs for each AP. However, our problem is different
  from the variants of group testing problems studied to date in the
  following aspects. In our setting, each test corresponds to a
  transmission pattern in the network, which is subject to constraints
  imposed by the direct interference graph, and depends on the traffic
  statuses of all APs. Therefore, we do not get to design the test
  matrix in the passive interference estimation setting.  This is in
  contrast to other group testing problems wherein the design of test
  matrices can be controlled. Moreover, the complicated transmission
  mechanism of the APs imposes a heavily constrained structure on the
  design matrix, which makes existing detection methods and analyses
  inapplicable. In addition, to recover of the hidden interference
  graph we are required to perform a number of overlapped group tests
  for all APs in parallel.  This results in quite different scaling
  laws of the number of tests required.

While we note that the motivating work for this paper comes from
  research into WLANs, we anticipate that these ideas will also find
  application in larger-scale cellular-type networks.  In particular,
  consider the heterogeneous networks (HetNets) that are part of the
  fourth-generation Long Term Evolution (LTE) Advanced networks. In
  HetNets, small ``femto'' cells interfere with large ``macro'' cells.
  Due to the overhead required, it is preferable to discover the
  interference environment using passive methods, rather than active
  probing methods that consume spectral
  resources~\cite{LiuEtAl:13,Lin:2015:OUA}. While HetNets are part of
  LTE-Advanced, they are anticipated to play an even more central role
  in fifth generation (5G) networks.  Further, in 5G much more
  back-end coordination between base stations is anticipated in the
  guise of Cloud Radio Access Networks or
  C-RANs~\cite{Simeone:16}. Such back-end coordination is just the
  mechanism needed to assemble the passively sensed data that we
  require to estimate the network's interference environment.

We adopt the following set of notations.  We use upper-case, e.g.,
$X$, and bold-face upper case, e.g., $\Xv$, to denote random variables
and vectors, respectively. A vector without subscript consists of $n$
elements, each corresponds to an AP, e.g.,
$\Xv:=(X_1,X_2,\ldots,X_n)$. Sets and events are denoted with
calligraphic font (e.g., $\Ec$). The cardinality of a finite set $\Vc$
is denoted as $|\Vc|$. A vector with a set as its subscript consists
of the elements corresponding to the transmitters in the set, e.g.,
$\Qv_{\Cc}:=\{Q_c\}_{c \in \Cc}$.

The paper is organized as follows. In Section~\ref{sec.formulation} we
formulate the WLAN interference identification problem as a graph
learning problem. We review the CSMA/CA (carrier sense multiple access
with collision avoidance) protocol and propose a statistical model for
a network using this protocol. In Section~\ref{sec:main} we present
our main results in the form of matching upper and lower bounds (up to
constant factors) on the observation requirements for reliable
estimation of both direct and hidden interference graph. We present an
experimental study in Section~\ref{sec.simResults} that supports the
theoretical analysis.  The experiments are based on simulations of the
traffic and protocol that {incorporate more real-world effects} than
the models used to develop the theory. However, in the experiments the
scaling behavior observed does match that predicted by theory.
Concluding remarks are made in Section~\ref{sec.conclusions}.  Most
proofs are deferred to the appendices.

\section{Problem Formulation}
\label{sec.formulation}

In this section we present the problem setting.  In
Section~\ref{sec.CSMA_CA} we present the important characteristics of
the multiple-access protocol.  In Section~\ref{sec.graphModel} we
model the interference environment using a graph.  In
Section~\ref{sec.interfAlg} we present the estimation problem and
sketch our algorithms. Finally, in Section~\ref{sec:model} we present
the statistical model and assumptions that underlie our analysis.

\subsection{CSMA/CA protocol and ACK/NACK mechanism}
\label{sec.CSMA_CA}

We assume the wireless networks operates using a CSMA/CA-like protocol
at the medium access control layer, e.g.,\cite{ieee802}. The important
characteristics of the protocol are as follows.  When a node has a
packet to send, it listens to the desired channel. If the channel is
idle, it sends the packet. If the channel is busy, i.e., there exists
an active transmitter within the listener's carrier sensing range, the
node waits for the end of that transmission, and then starts to
contend for the channel. To contend, the node randomly chooses an
integer $w$, uniformly distributed in the range $[0,W-1]$, and then
the node backs off for $w\times \tau$ seconds. The positive integer
$W$ represents the back-off window size, and $\tau$ is the duration of
a time slot. If the channel is idle at the end of the node's back-off
period, it transmits its packet. The node that chooses the smallest
back-off time wins the channel and transmits its packet. The other
nodes wait for the next round of contention at the end of the
transmission of this packet. According to this protocol, roughly
  speaking, the time axis can be partitioned into sessions, where each
  session consists of a contention period followed by a transmission
  period.  Statistically, the random back-off mechanism allows every
node equal access to the medium.

However, even if two APs are not within each other's carrier sensing
range, the transmission from one AP may still corrupt the signal
received at clients of the other.  This is the so-called ``hidden
terminal'' problem. To identify this type of interference, additional
information is needed.  We assume that an ACK/NACK mechanism is
used. Specifically, we assume that whenever an AP successfully
delivers a packet to its destination, an ACK is fed back to
acknowledge the successful transmission. If the AP does not receive
the ACK after a period of time, it assumes that the packet was
lost. The ACK/NACK mechanism enables the APs to detect collisions
however they may occur.  Throughout we assume that all ACKs are
reliably received at the corresponding transmitters.

\subsection{Interference Graph}
\label{sec.graphModel}

\begin{figure}
\centerline{ \includegraphics[width=3in]{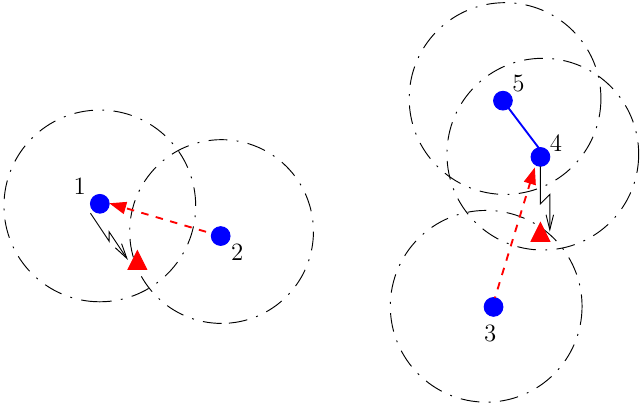}}
\caption{System model with $5$ APs. The edges connecting APs represent
  the interference between them. Solid line represents direct
  interference, dashed arrows represent hidden interference. Each
  circle represents the carrier sensing range of the AP at its center.
  Clients (represented by triangles) that are located in the intersection of the carrier sensing
  regions of a pair of APs may be subject to interference.}
\label{fig:int_graph}
\end{figure}

We use a graph $G=(\mathcal{V},\mathcal{E})$ to represent the
interference among APs in the network.  The node set $\mathcal{V}$
represents the APs, and the edge set $\mathcal{E}$ represents the
pairwise interference among APs. An example of such a graph is
depicted in Figure~\ref{fig:int_graph}.

We partition $\mathcal{E}$ into two subsets: {\em direct} interference
$\mathcal{E}_D$ and {\em hidden} interference $\mathcal{E}_H$. Direct
interference occurs when two APs are within each other's carrier
sensing range. {An example in Figure~\ref{fig:int_graph} is the AP
  pair $(4,5)$.}  Under the assumption that the carrier sensing range
is the same for every AP, the edges in $\mathcal{E}_D$ are {\it
  undirected}. However, as mentioned earlier, carrier sensing cannot
resolve all of the collisions in the network.  Hidden-terminal type
interference is represented by the edges in $\Ec_H$.  Such
interference may be asymmetric and so the edges in $\Ec_H$ are {\it
  directed}.  {In Figure~\ref{fig:int_graph} AP pairs $(1,2)$ and
  $(3,4)$ cannot detect each other.  Yet, they can interfere with each
  other's clients since their carrier sensing ranges intersect.  The
  affected clients would (roughly) lie in the intersection of the
  (roughly circular) carrier sensing ranges of the AP pair.  Nodes $3$
  and $5$ is an example of a pair of APs that do not interfere with
  each other (either directly or indirectly) since their carrier
  sensing ranges do not overlap.}

\subsection{Inference Problem and Algorithm}
\label{sec.interfAlg}

Our objective is to determine the edge set $\mathcal{E}$ based on
passive observations. We assume the existence of a central controller
that, over a period of time, collects the transmission statuses of all
APs and information on transmission successes (ACKs). The collected
information during this {\it observation period} is the dataset we use
to estimate the interference graph.  Both enterprise-type WLANs and
proposed architectures for 4G systems that involve combinations of
macro plus pico-cells have such central controllers.

In this paper we consider {\it static
  channel states}. Specifically, we assume that whether or not a pair
of APs can sense each other is fixed (deterministic) throughout the
period of observation. {For hidden interference, we also assume
  that the channel between each AP and all clients is
  also static.  Further, if a transmission failure occurs it must have
  been caused by a collision with a packet transmitted by one of the
  other active APs.

We now outline the basic ideas underlying our algorithms.  First,
consider the direct interference environment, characterized by
$\mathcal{E}_D$.  Note that due to the use of carrier sensing in
CSMA/CA, if two APs are within each other's carrier sensing range,
they will not transmit simultaneously. One can therefore infer that
any pair of APs that transmit simultaneously during the observation
period must not be able to hear each other.  In other words, there is
no {\it direct} interference between them.  The algorithm starts with
a full set of ${{|\mathcal{V}|}\choose{2}}$ candidate edges.  Each
time a simultaneous transmission is observed, the corresponding edge
is removed.  If the observation period is sufficiently long, all
possible co-occurring transmission will be observed and the correct
{\it direct} interference edge set $\Ec_D$ will be recovered.

Next, consider the cases of hidden interference characterized by
$\mathcal{E}_H$.  Estimating this set is more involved and requires
the collected ACK information. When a collision is detected at an AP,
it implies that at least one of the other APs transmitting at the same
time is interfering. The subset of APs transmitting at that time
instance is a candidate set of hidden interferes for that AP.  For
each collision detected by the AP, another candidate subset is
formed. The set of hidden interferers for that AP must have a
non-empty intersection with {\em all} such candidate sets. When the
observation period is sufficiently long, we show that the {\it minimum
  hitting set} \cite{karp} that intersects with all of the
candidate subsets is the set of hidden interferers. The edge set
$\Ec_H$ can thus be recovered.

Our approaches to both problems were inspired by the PIE algorithms
proposed in \cite{vivek_thesis,pie_nsdi}. For the direct interference
problem, our approach is similar to PIE.  Both rely on simultaneous
transmissions to infer the interference graph.  Our results provide
the theoretical analysis and scaling behavior characterization that
complement the empirical results of~\cite{vivek_thesis,pie_nsdi}. Our
approach to the hidden interferences problem is quite different from
that taken in~\cite{vivek_thesis,pie_nsdi}.  While PIE uses a
correlation based approach we use the hitting set approach described
above.  The hitting set approach results in more accurate estimation
of the graph of hidden interferers, as elaborated in
Sec. \ref{sec.achHiddenInt}.

\subsection{Statistical Model}
\label{sec:model}

The graph $G_D=(\Vc,\Ec_D)$ represents the carrier sensing
relationships among the APs. Specifically, if AP $i$ and AP $j$ are
within each other's carrier sensing range, there is an edge between
$i,j$, denoted as $(i,j)$. The existence of such an edge implies that
AP $i$ and AP $j$ are close and the transmission of each can be
observed by the other.  An example in Figure~\ref{fig:int_graph} is
the AP pair $(4,5)$. We term this ``direct'' interference and $G_D$
the direct interference graph. We define $\mathcal{N}_i$ to be the set
of neighbors of AP $i$ in $G_D=(\Vc,\Ec_D)$, and let
$d_i=|\mathcal{N}_i|$.

Observations of the network activation pattern are taken each time
epoch. We use $X_i(t) \in \{0,1\}$ to denote the activation state of
node $i$ at time $t$: $X_i(t)=1$ means that node $i$ transmits;
$X_i(t)=0$ means that node $i$ does not transmit. In general, $X_i(t)$
is determined by the traffic statuses and backoff times of the APs
that compete for the same channel. We denote the traffic status 
of node $i$ as $Q_i(t)$. $Q_i(t)$ is a Bernoulli random variable, and it equals
one if node $i$ has some content to send in slot $t$; otherwise, it equals zero.

As mentioned, to infer the hidden interference we require information
on transmission successes and failures.  Define
$Y_i(t)\in\{0,1,\varnothing\}$ to be the feedback information received
at AP $i$ at the end of session $t$. $Y_i(t)=1$ means that an ACK is
received at AP $i$, indicating that the transmission in session $t$ is
successful; $Y_i(t)=0$ means that the transmission has failed, caused
by some simultaneous transmission(s); $Y_i(t)=\varnothing$ means that
node $i$ did not transmit in that session, i.e., $X_i(t)=0$.

The graph $G_H=(\Vc,\Ec_H)$ represents the hidden interference among
APs that cannot hear each other. This interference depends on the
locations of the clients associated with each AP and thus may not be
symmetric. These edges are therefore {\it directed}. We define
\begin{align}\label{eqn:pij}
p_{ij}&=\Pd(Y_j(t)=0|X_i(t)=X_j(t)=1, \Xv_{\Vc\backslash\{i,j\}}(t)=\ov)
\end{align}
{i.e., $p_{ij}$ is the probability that, when $i,j$ are isolated from
  the rest of the APs and $X_i(t)=X_j(t)=1$, AP $i$ interferes with AP
  $j$, causing transmission failure of AP $j$. }

The $p_{ij}$ capture the randomness in locations of the clients
associated with each AP.  An AP may interfere with only a subset of
the clients of a neighboring AP.  For example, in
Fig.~\ref{fig:int_graph} AP $1$ may interfere with a client of AP $2$
that is halfway between the APs, but it likely will not interfere with
a client on the far side of AP $2$.  Thus, which clients of an AP
another AP interferes with depends on the locations of those
clients. The value of $p_{ij}$ represents the proportion of AP $j$'s
clients that AP $i$ interferes with.  It can be interpreted as the
probability that AP $j$ communicates with a client located in the
overlapped area of the carrier sensing ranges of APs $i$ and $j$. Note
that $p_{ij}$ is not defined for $(i,j)\in \Ec_D$ since such pairs of
APs are within each other's carrier sensing range and thus never
transmit simultaneously. This is the case for AP 4 and AP 5 in
Fig.~\ref{fig:int_graph}.

Define $\Sc_j = \{ (i,j) \in \mathcal{E}_H \ | \ i \in \mathcal{V}\}$
to be the hidden interferer set for AP $j$, i.e., the set of APs with
$p_{ij}> 0$. We let $s_j=|\Sc_j|$.  We point out that in general
$\forall \Sc\subseteq \Vc\backslash\{i,j\}$,
 \begin{align}\label{eqn:pij_bound}
 p_{ij}\leq \Pd(Y_j(t)=0|X_i(t)=1,X_j(t)=1,\Xv_{\Sc}=\ov),
\end{align}
i.e., $p_{ij}$ is {\em less} than the probability that a collision
occurs at AP $j$ when both APs $i$ and $j$ are transmitting. This is
because the collision at AP $j$ may be caused by an active AP other
than $i$, and AP $i$ may just happen to be transmitting at the same
time.

The complete interference graph $G=(\Vc,\Ec)$ consists of both direct
interference graph and hidden interference graph, i.e.,
$\Ec=\Ec_D\cup\Ec_H$. We note that $\Ec_D\cap\Ec_H=\varnothing$.

While we assume pairwise interference in our system model, the
  interference graph definition can be generalized to include more
  than just pairwise relationships. If certain types of interference
  can occur only when multiple APs are simultaneously transmitting,
  those APs form a virtual ``super'' interferer. We can include all
  possible super interferers as nodes in the inference graph without
  fundamentally changing the interference process.

We now state some statistical assumptions important in the development
of our analytic results. 

\begin{Assumption} \label{firstAssump} \hspace{1em}
\begin{enumerate}
\item[(0)] APs are synchronized and the time axis is partitioned into
  synchronized sessions. We use $t=1,2,\ldots$ to denote the indices
  of the observation sessions.
\item[(i)] The traffic status $Q_i(t)$ are i.i.d.\ Bernoulli random
  variables with common parameter $p$, where $0<p<1$, for all $i \in
  \mathcal{V}$ and all $t \in \mathbb{Z}^+$.  In other words, the
  $Q_i(t)$ are independent across transmitters and sessions:
  $\Pd(Q_i(t)=1)=p$ for all $i$ and $t$. APs competes for the channel
  at the beginning of session $t$ if and only if $Q_i(t)=1$.
\item[(ii)] In the contention period at the beginning of each session,
  the backoff time $T_i(t)$ for competing APs are continuous
  i.i.d. random variables uniformly distributed over $[0,(W-1)*\tau]$
  and are statistically independent for all $i \in \mathcal{V}$ and
  all $t \in \mathbb{Z}^+$. $W$ is a fixed positive integer, and
  $\tau$ is the duration of a time slot.
\item[(iii)] For all $(i,j) \in \mathcal{E}_H$, there exists a
  constant $p_{min}$, $0<p_{\min}<1$, s.t. $p_{ij} \geq p_{\min}$.
\item[(iv)] For all $i \in \mathcal{V}$, there exists an integer $d\in
  \mathbb{Z}^+$, s.t. $d_i \leq d$, i.e., the number of direct
  interferers of any AP is upper bounded by $d$.
\item[(v)] For all $j \in \mathcal{V}$, there exists $s\in
  \mathbb{Z}^+$, s.t. $s_j \leq s$, i.e., the number of hidden
  interferers of any AP is upper bounded by $s$.
\end{enumerate}
\end{Assumption}

In the following, we use $\mathcal{G}_d$ to denote the set of
direct interference graphs consisting of $n$ nodes and satisfying
Assumption~\ref{firstAssump}-(iv). Since whether an AP is a
  hidden interferer with respect to other APs depend on whether or not
  they can hear each other, the hidden inference graph of a network
  actually depend on its direct interference graph. Thus, for any
given $G_D\in \mathcal{G}_d$, we define $\mathcal{H}_s(G_D)$ as
the set of hidden interference graphs satisfying
Assumption~\ref{firstAssump}-(v).

Assumption~\ref{firstAssump}-(0) is essential for our analysis. In
practical WLANs, APs competing for the same channel are ``locally''
synchronized because of the CSMA/CA protocol.
Assumption~\ref{firstAssump}-(0) approximates the original ``locally''
synchronous system as a synchronized one. Since APs far apart on the
interference graph have a relatively light influence on each other's
activation status, this assumption provides a good approximation of
the original system and greatly simplifies our analysis. 

Assumption~\ref{firstAssump}-(i) ignores the time dependency and
coupling effect of the traffic queue statuses of APs. This assumption
is made to simplify the analysis. In a practical setting, the
traffic queue status for a single AP is coupled from slot to slot,
depending on whether the AP got access to the channel in the previous
slot. The queue statuses of different APs can also be coupled
(directly or indirectly) due to the common channels for which they
compete. However, as long as the queues in the network are
  stable, their mixing time is finite from a queuing theory
  perspective.  We can therefore always sample the network every $T$
  sessions so that the queue status dependency between two consecutive
  observation sessions are negligible when $T$ is sufficiently
  large. This would add only a constant factor $T$ to the required
  observation duration. However, we note that such an addition does not
  change our scaling results. Also, as indicated by our simulation
  results in Sec.~\ref{sec.simResults}, our algorithms do not require
  Assumption~\ref{firstAssump}-(i) in order to function properly, so
  subsampling is, in fact, unnecessary.

The purpose of Assumption~\ref{firstAssump}-(ii) is to ensure
  that with probability one no two adjacent nodes in $G_D=(\Vc,\Ec_D)$
  have the same back-off time and therefore transmission collisions
  are avoided completely. Although in the CSMA/CA protocol, the
  backoff times are integer multiples of $\tau$, this assumption is a
  reasonable approximation when $W$ is a large integer. Roughly
  speaking, the probability that two adjacent nodes in $G_D$ choose
  the same backoff time under the CSMA/CA protocol is upper bounded by
  $1/W^2$.  Given that the number of edges in $G_D$ is upper bounded
  by $nd/2$, and through application of the union bound, the
  probability that a collision happens is upper bounded by
  $\frac{nd}{2W^2}$. Further, asynchrony of AP operations in real
  networks make the probability of collision even smaller.  We thus
  make Assumption 1-(ii) to simplify our analysis without compromising
  much in terms of accuracy. Finally, we note that this assumption is
  applied only for analytical purposes and is not applied in the
  experimental section where we observe the predicted scaling laws
  under a more realistic simulation of protocol operations. We comment
  that we did analyze a model wherein collisions are allowed to occur,
  and we did extend our interference estimation algorithm to handle
  this case.  We are not able to include those results herein due to
  space constraints, but they can be seen in the online version of the
  paper \cite{int_graph_arxiv}.

The first two assumptions guarantee that the joint distribution of
$X_i(t)$s and $Y_i(t)$s is independent and identical across
$t$. Therefore, in the analysis hereafter we ignore the time index and
focus on the distribution of $X_i$s and $Y_i$s in a single
observation.

Regarding the last three assumptions,
assumption~\ref{firstAssump}-(iii) defines a lower bound on the level
of interference of interest.  The final assumptions, (iv) and (v),
model the fact that the interference graph will be sparse because of
the widespread spatial distributions natural to the large-scale
wireless networks of interest.

\section{Main Results}
\label{sec:main}

In this section we present our main results.  We break the overall
problem into four subproblems. In Section~\ref{sec.achDirectInt} we
provide an achievable upper bound on the number of observations
required to infer the directed interference edge set $\Ec_D$.  In
Section~\ref{sec.convDirectInt} we present a matching lower bound on
the number of observations required to infer $\Ec_D$. In
Section~\ref{sec.achHiddenInt} we provide an achievable upper bound on
the number of observations required to infer the hidden interference
edge set $\Ec_H$. Finally, in Section~\ref{sec.convHiddenInt} we
present a lower bound on the number of observations required to infer
$\Ec_H$.

\subsection{An Upper Bound for Determining $G_D=(\Vc,\Ec_D)$ }
\label{sec.achDirectInt}

In this section, we formalize the algorithm sketched in
Section~\ref{sec.interfAlg} for estimating $G_D=(\Vc,\Ec_D)$ and
present an analysis thereof. In the static setting, for any
$(i,j)\in\Ec_D$, APs $i$ and $j$ can always sense each other's
transmissions.  For any AP pair $(i,j)\notin \Ec_D$, the APs can never
detect each other's transmissions.

Say that a sequence of transmission patterns $\Xv(1)$, $\Xv(2)$,
$\ldots$, $\Xv(k)$ is observed. Due to the use of the CSMA/CA protocol
and the continuous back-off time of Assumption~\ref{firstAssump}-(ii),
any pair of APs active in the same slot must not be able to hear each
other. Thus, there is no edge in $G_D=(\Vc,\Ec_D)$. In other words,
given an observation $\Xv$, for any $i,j$ with $X_i=X_j=1$, we know
that $(i,j)\notin \Ec_D$.

Based on this observation, the algorithm starts at $t=1$ with a fully
connected graph connecting the $n$ APs with
${{|\mathcal{V}|}\choose{2}} = n(n-1)/2$ edges. For each transmission
pattern $\Xv$ observed, we remove all edges $(i,j)$
s.t. $X_i=X_j=1$. Our first result quantifies the number of
observations $k$ required to eliminate, with high probability, all
edges not in $\Ec_D$, thereby recovering the underlying interference
graph $G_D$.  In Appendix~\ref{apx:upper1} we show the following
result:
\begin{Theorem}\label{thm:upper}
Let $\delta>0$, and let
\begin{align}\label{eqn:upper}
k&\geq \frac{1}{\log \frac{1}{1-p^2/(d+1)^2}}\left(\log{n \choose 2}
+\log \frac{1}{\delta}\right).
\end{align}
Then, with probability at least $1-\delta$, the estimated interference
graph $\hat{G}_D=(\Vc,\hat{\Ec}_D)$ is equal to $G_D$ for any $G_D\in \mathcal{G}_d$ after $k$
observations.
\end{Theorem}

The idea of the proof is first to lower bound the probability that two
nonadjacent APs $i,j$ {\em never} transmit simultaneously in $k$
observations. Then, by taking a union bound, an upper bound on the
required $k$ is obtained.

\textbf{Remark:} If $p^2/ d^2 \ll 1$ the scaling on $k$ in the
  theorem simplifies to $O(d^2\log n)$.  We see this by noting that if
  $p^2 / d^2 \ll 1$ then $-\log \left(1-p^2/(d+1)^2\right)$ is well
  approximated by $p^2/d^2$ and $\log{n \choose 2}$ scales as $\log n$.
  Most networks of interest will fall in this regime, e.g., if $p=0.5$
  and $d=3$ then $p^2/d^2$ will be a good approximation of $-\log
  \left(1-p^2/(d+1)^2\right)$. This sort of scaling in $p$ and $d$ can
  easily be believed to be the best we can hope for when passive
  estimation is employed.  This follows because if two non-interfering
  APs never transmit simultaneously, their behavior is the same as if
  they were within each other's carrier sensing range. Thus, we cannot
  determine whether or not there is an edge between them. Since each
  transmitter competes with its neighbors, the probability that it
  gets the channel is roughly $p/d$. Thus, it takes about $d^2/p^2$
  snapshots to observe two non-interfering APs active at the same
  time. Since there are about $n^2$ such pairs in the network, an
  application of the union bound yields the factor $\log n$.

\subsection{A Minimax Lower Bound for Determining $G_D=(\Vc,\Ec_D)$ }
\label{sec.convDirectInt}

We now provide a minimax lower bound on the number of observations
needed to recover the direct interference graph
$G_D=(\Vc,\Ec_D)$. Denoting the estimated graph as
$\hat{G}_D=(\Vc,\hat{\Ec}_D)$ we prove the following result
in~Appendix~\ref{apx1}.

\begin{Theorem}\label{thm:low1}
For any $\alpha$, $0 < \alpha < 1/8$, if $n \geq 7$, $2\leq d\leq (3n-\sqrt{n^2+16n})/4$, and
\begin{equation*}
k\leq \frac{ \alpha d^2}{\left(2+\frac{1}{1-p}\right)}\log n,
\end{equation*}
then,
\begin{align}\label{eqn:minmax}
\min_{\hat{G}_D\in \mathcal{G}_d}\max_{G_D\in \mathcal{G}_d} & \Pd(\hat{G}_D\neq
G_D;G_D)\nonumber\\
& \geq \frac{\sqrt{n}}{1+\sqrt{n}}\left(1-2\alpha-\sqrt{\frac{2\alpha}{\log
    n}}\right).
\end{align}
\end{Theorem}

The approach to deriving this result is as follows.  We construct a
set of $M$ maximum-degree $d$ graphs. We construct the set so that the
graphs in the set are very similar to each other.  This makes it hard
to distinguish between them.  For each graph in the set the
statistical assumptions of Section~\ref{sec:model} induce a
distribution on the observed transmission patterns.  Given $k$
observed transmission patterns we consider the $M$-ary hypothesis test
to detect the underlying graph.  {As the size of the candidate set for
  the original estimation problem is much greater than $M$,} this test
will be easier than the original problem. Therefore, a lower bound for
this test will also lower bound the original estimation problem.
Since for each graph we know the distribution of patterns, we can
lower bound the probability of error for this hypothesis test using
the Kullback-Leibler divergence between each pair of induced
distributions.

\textbf{Remark:} We note that the RHS of (\ref{eqn:minmax}) is
  bounded away from zero, and $\frac{\sqrt{n}}{1+\sqrt{n}}\geq
  \frac{\sqrt{3}}{1+\sqrt{3}}$ for $n\geq 3$. Therefore, for any
  $\delta<\frac{\sqrt{3}}{1+\sqrt{3}}$, we can always find a positive
  $\alpha$ so that $\min_{\hat{G}_D}\max_{G_D}\Pd(\hat{G}_D\neq
  G_D;G_D)\geq \delta$. Then, the number of observations required to
  detect the correct underlying graph with probability $1-\delta$ is
  $\Omega(d^2\log n)$ which, as discussed earlier, when $p^2/d^2 \ll
  1$, is the same order as the upper bound in
  Theorem~\ref{thm:upper}. Therefore, the estimation method based on
  pairwise comparison is asymptotically optimal when $d$ is large.

\subsection{An Upper Bound for Determining  $G_H=(\Vc,\Ec_H)$}
\label{sec.achHiddenInt}

We now present our results on inferring the hidden interference
graph $G_H=(\Vc,\Ec_H)$.  We observe $\Xv(1), \ldots \Xv(k)$ and
$\Yv(1), \ldots \Yv(k)$.  When the transmission of AP $j$ fails
(indicated by the feedback $Y_j = 0$) the failure must have been
caused by collision with a transmission from one of the active APs in
$\Sc_j$.  However, as there may be multiple hidden interferers, there
may be no single AP that is always transmitting when
$Y_j=0$. Complicating the situation is the fact that an AP that
transmits regularly when $Y_j=0$ may not be a hidden interferer at
all. This is because the transmission status of an AP can be strongly
positively correlated (because of CSMA/CA) with one or more non-interferers.  We first illustrate the types of statistical
dependencies we need to address before presenting our algorithm.

First, consider a scenario where AP $1$, AP $2$, and AP $3$ lie in the
direct interference range of AP $l$, i.e., APs $1$, $2$, and $3$ are
all in $\Nc_l$.  Further assume that there is no direct interference
between $1$, $2$, and $3$.  The transmission of any of these three APs
will suppress the transmission of AP $l$ and thus increase the
(conditional) probability of transmission of the other APs in $\Nc_l$.
Thus, the activation statuses of $1,2,3$ are positively
correlated. Now say that APs $1,2$ are both hidden interferers for AP
$m$, $m\notin\{1,2,3,l\}$.  Then, even though transmissions of AP $3$
may be correlated with transmission {\em failures} of AP $m$, due to
the positive correlation statuses of $1,2,3$, AP $3$ may not be a
hidden interferer for node $m$.  {One possible scenario is that
  $\Pd(Y_m=0|X_3=1,X_m=1)$ may be even greater than
  $\Pd(Y_m=0|X_1=1,X_m=1)$ or $\Pd(Y_m=0|X_2=1,X_m=1)$.} The upshot is
that correlation-based approaches to determining hidden interferers,
such those adopted in~\cite{vivek_thesis,pie_nsdi}, may not be able
distinguish true interferers from non-interferers. To address these
issues we propose the following approach, based on {\em minimum
  hitting sets}.

First, given $k$ observations, define
\begin{equation*}
\Kc_j(k)= \{t \in \{1, 2, \ldots, k\} \ | \ Y_j(t)=0\}
\end{equation*}
to be the sessions in which AP $j$'s transmissions fail.  For each $j$
and each $t$ define the set of candidate hidden interferers as
\begin{equation*}
\Sc^t_j = \{ i \in \mathcal{V} \ | \ i \neq j, X_i(t) = 1\}.
\end{equation*}
We will be interested only in $t$ such that $t \in \Kc_j(k)$.  Our
estimator of the set of hidden interferers $\hat{\Sc}_j(k)$ is the
{\em minimum hitting set} of the candidate interferer sets
$\{\Sc_j^t\}_{t\in \Kc_j(k)}$:
\begin{equation*}
\hat{\Sc}_j(k) = \arg\min_{\Sc\subseteq \Vc}\{ |\Sc| \ |
\ \Sc\cap\Sc_j^t\neq \varnothing, \forall \ t\in \Kc_j(k)\},
\end{equation*}
where, if there are multiple minimizers, one is selected at random.

In general, a minimum hitting set is defined as follows:
\begin{Definition} {\em \bf (Minimum Hitting Set)}
  Given a collection of subsets of some alphabet, a set which
  intersects all subsets in the collection in at least one element is
  called a ``hitting set''. A ``minimum'' hitting set is a hitting set
  of the smallest size.
\end{Definition}

Given $k$ observations, our algorithm determines the minimum hitting
set $\hat{\Sc}_j(k)$ of $\{\Sc^t_j \ | \ t \in \Kc_j(k)\}$ for each $j
\in \mathcal{V}$. To get a sense of the usefulness of the concept
  of the minimum hitting set when considering hidden interferers,
  consider when a single AP $j$ has two hidden interferers, both of
  which are active for all $t \in \Kc_j(k)$.  Since we look for the
  {\em minimum} hitting set only one of these would be included in the
  graph estimate.  However, if we wait sufficiently long, we will
  experience a pair of transmissions in which each of these two hidden
  interferers is solely active.  At that point both will be included
  in the graph estimate.  In contrast, while if we don't require the
  {\em minimum} hitting set to be our estimate both hidden interferers
  might be included in out estimated graph earlier, so might be other
  nodes that are non-interfering but happen always to be active in
  conjunction with various distinct hidden interferers.

The following theorem provides an upper bound on the number of
observations required so that $\hat{\Sc}_j(k) = \Sc_j$ for every AP
$j\in\Vc$ with high probability. Once the estimated minimum hitting
set $\hat{\Sc}_j(k)$ is obtained, the estimated hidden interference
graph $\hat{G}_H$ is constructed by adding a directed edge from each
AP in $\hat{\Sc}_j(k)$ to AP $j$, i.e.,
\begin{equation*}
\hat{\mathcal{E}}_H = \bigcup_{j \in \mathcal{V}} \left\{(i,j) \ | \ i
\in \hat{\Sc}_j(k)\right\}.
\end{equation*}
The following theorem is proved in Appendix~\ref{apx:weak_upper}.

\begin{Theorem}\label{thm:weak_upper}
Let $\delta>0$, and let
\begin{align*}
k&\geq  \frac{1}{\log \frac{1}{1-\frac{p^2(1-p)^{s}p_{\min}}{(d+1)^2}}}\left(\log (ns) +\log \frac{1}{\delta}\right)
\end{align*}
Then, with probability at least $1-\delta$, $\hat{G}_H$ equals $G_H$ for any given $G_D\in \mathcal{G}_d$ and $G_H\in \mathcal{H}_s(G_D)$.
\end{Theorem}

The approach taken in the proof can be summarized as follows.  For
every AP $j$, we first upper bound the probability that the minimum
hitting set obtained after $k$ observations is not equal to the true
minimum hitting set. This is equal to the probability that there
exists at least one AP $i \in \Sc_j$ that is not included in
$\hat{\Sc}_j$. As mentioned above, this will happen (at least) if
  two hidden interferers happen to both be on in all $t \in
  \Kc_j(k)$.  By taking the union bound across all possible $i$ and
$j$, we obtain an upper bound on $k$. The upper bound
$k=O\left(\frac{d^2}{p^2(1-p)^sp_{min}}\log n\right)$ when
$\frac{p^2(1-p)^sp_{min}}{d^2}\ll 1$.


In general, finding the minimum hitting set is NP-hard
\cite{karp}. However, under the assumption that the maximum
  possible number of hidden interferers $s \ll n$, the total
  number of nodes in the network, the minimum hitting set can be
solved for in polynomial time.  This is a regime appropriate to the
large-scale wireless networks of interest in this paper.  First we use
the algorithm of Section~\ref{sec.achDirectInt} to identify $G_D$.
Next, we consider each AP in turn.  For AP $j$, we test every subset
of nonadjacent APs in $G_D$ (not including AP $j$) to determine
whether it is a hitting set of $\{\Sc^t_j \ | \ t \in
\Kc_j(k)\}$. Since the number of hidden interferers $s_j \leq s$, we
start with the smallest possible hitting sets, i.e., $s_j = 1$.  We
increment the size of the testing subset by one, until a hitting set
is achieved. In this way we find the minimum hitting set for the given
$\{\Sc^t_j \ | \ t \in \Kc_j(k)\}$.

The worst situation for this incremental approach will be when $s_j$
is as large as possible.  By Assumption~\ref{firstAssump}-(v) $s_j
\leq s$.  Recalling that $d_j$ is the number of direct interferers
(which can be eliminated from consideration), in this case the maximum
number of subsets we must test is
\begin{align*}
\sum_{i=1}^s{n-d_j-1 \choose i}=\beta_3 (n-d_j-1)^s
\end{align*}
for some bounded constant $\beta_3$.  In other words, the number of
subsets we need to test is upper bounded by $O(n^s)$ \cite{pc_dag}.

\subsection{A Minimax Lower Bound for Determining $G_H=(\Vc,\Ec_H)$}
\label{sec.convHiddenInt}

Finally, we provide a lower bound on the number of observations
required to recover the underlying hidden interference graph
$G_H=(\Vc,\Ec_H)$.  In Appendix~\ref{apx:low3} we prove the following
theorem.

\begin{Theorem}\label{thm:low3}
 Assume $s\geq 2$. For any $c_1,c_2>0$ s.t.
\begin{equation}
d+1 \leq c_1 n, \ \ \
s-1  \leq c_2 n,  \ \ \ 2c_1+c_2<1. \label{eq.scalingCond}
\end{equation}
If
\begin{equation*}
k\leq \frac{\log\frac{1}{
   2c_1(\frac{1}{2c_1+c_2}-1)n}}{\left(\frac{1-(1-p)^{d+1}}{d+1}\right)^2(1-p)^{s-1}\log
   (1-p_{\min})}
\end{equation*}
then for any $\alpha$, $0 < \alpha < 1/8$,
\begin{align*}
 &\min_{\hat{G}_H\in\mathcal{H}_s(G_D)}\max_{G\in \mathcal{G}_d\times \mathcal{H}_s(G_D)}\Pd(\hat{G}_H\neq
  G_H;G) \geq\\ &\hspace{-0.03in}\frac{\sqrt{2c_1(\frac{1}{2c_1+c_2}-1)n}}{\!1\hspace{-0.03in}+\hspace{-0.03in}\sqrt{2c_1(\frac{1}{2c_1+c_2}-1)n}}\left(1-\hspace{-0.03in}2\alpha\hspace{-0.03in}-\hspace{-0.03in}\sqrt{\frac{2\alpha}{\log
      (2c_1(\frac{1}{2c_1+c_2}\hspace{-0.03in}-\hspace{-0.03in}1)n)}}\right)\!.
 \end{align*}
\end{Theorem}

The first two conditions expressed in~(\ref{eq.scalingCond}) ensure
that the maximum number of direct interferers $d$, and the bound on
the number of hidden interferers $s$, both scale at most linearly in
$n$.  The third constraint places a limit on the joint scaling.  The
approach to deriving this result is the same as the one we followed
for the proof of Theorem~\ref{thm:low1}.  We reduce the original
problem to an $M$-ary hypothesis test and show that, asymptotically,
the lower bound has the same order as the upper bound.

\textbf{Remark:} Since the distribution of $\Yv$ depends on the
underlying direct interference graph as well as on the hidden
interference graph, the lower bound is over all possible interference
graphs $G$. As we discussed after Theorem~\ref{thm:low1}, for a
  sufficiently small positive number $\delta$, we can always find a
  positive $\alpha$ so that
  $\min_{\hat{G}_D}\max_{G_D}\Pd(\hat{G}_D\neq G_D;G_D)\geq
  \delta$. Then, as $d$ increases, the number of observations required
  to detect the correct underlying graph with probability $1-\delta$
  is $\Omega\left(-\frac{d^2}{\log(1-p_{\min})(1-p)^{s-1}}\log
  n\right)$. Since $\log (1-p_{\min})$ can be approximated as
$-p_{\min}$ when $p_{\min}$ is small, the lower bound is of the same
order as the upper bound provided in
Theorem~\ref{thm:weak_upper}. Therefore, the bounds are tight and our
hitting-set based algorithm is asymptotically optimal.

\section{Simulation Results}
\label{sec.simResults}

In this section we present simulation results with the aim of
  verifying the applicability of the theory developed earlier.  We use
  our algorithms to infer interference graphs based on traffic traces
  collected from a simulated wireless network that operates according
  to the IEEE 802.11 CSMA/CA protocol.  In comparison to the
  statistical assumptions made to derive the theoretical learning
  bounds in this paper, the
  simulations mimic real-world wireless networks much more closely.

We comment on the specific differences between the following
  simulations and the assumptions made in the analyses of
  Section~\ref{sec:main}.  First, the nodes in the simulations operate
  in an asynchronous fashion instead of working
  synchronously. Therefore, the synchronized session model of
  Assumption 1-(0) does not hold in the simulations. Second, we
  introduce a queue at the MAC layer for each AP to store data packets
  that haven't yet been delivered. A packet stays in the MAC queue
  until it has been successfully received at the destination client or
  has been dropped (after two retransmission attempts).
Therefore, the independent traffic status assumption made in
  Assumption 1-(i) does not hold. Third, the backoff time for each AP
  is not continuous but is an integer multiple of a slot time
  ($20\times 10^{-6}$ s). This means that two
  APs within each other's carrier sensing range have a (small)
  probability of colliding, especially when the window length $W$ is
  short. These differences break the i.i.d.~assumption regarding the
  joint distribution of $X_i(t)$ and $Y_i(t)$ across time. As we will
  see, despite the added complexities of this more realistic
  simulation environment, the behavior we observe closely matches the
  quantitative predictions made by the theory.

The specific setup for the simulations is as follows.  Access points
and clients are deployed over a rectangular area that can be
partitioned into square cells $50$m on a side.  {An AP is placed
  uniformly at random within each cell, while a client is placed at
  the center of the cell. Each client is associated with the nearest
  AP. We choose the network topology in
  this manner to ensure the randomness of the corresponding
  interference graph while still maintaining a relatively balanced
  traffic intensity across the network. Because there is a single
  client associated with each AP, in these simulations we are
  essentially evaluating the interference between AP-client links,
  similar to the setup in \cite{pie_nsdi}. } The locations of APs and
clients are fixed throughout the period of observation.

At the MAC layer, we generates an independent downlink traffic flow
for each client according to a Poisson process of $\lambda$ packet
arrivals per slot time. Packets payloads are all identical, of $1000$
bits each. {We set the contention window size to be 16 slot times.}

At the PHY layer, we employ the log-distance path loss model. In this
model, received power (in dB) at distance $l$ (in meters) from the
transmitter is given by:
\begin{align}\label{eqn:revPower}
\Gamma(l) &=\Gamma(l_0) -10\eta \log(l/l_0) + X_\sigma.
\end{align}
In the above, $\Gamma(l_0) $ is the signal strength at the reference
distance $l_0$ from the transmitter, $\eta$ is the path loss exponent,
and $X_\sigma$ represents a Gaussian random variable with zero mean
and variance $\sigma^2$ in dB. We choose $l_0$ to be 1 km, $\sigma^2$
to be 5dB, and $\eta$ to be 4. We also assume that the ``shadowing''
(represented by $X_\sigma$) between any AP and AP-client pair is fixed
throughout the period of observation. Thus the underlying interference
graph is constant within the period of observation.

We fix the transmission rate to be 5Mbps and the transmission range
for APs to be $37.5$m. The transmit power and corresponding
received SNR threshold are selected to ensure successful transmissions
within the transmission range.

We first study the direct interference estimation algorithm of
Section~\ref{sec.achDirectInt}. We fix the carrier sensing range for
the APs to be $60$m.  We vary the size of the network where the
network consists of an array of square cells. For each network size,
we randomly generate ten topologies, i.e., AP positions are randomly
chosen. For each of the ten topologies, we use our algorithm to
recover $G_D$ under different (randomly generated) traffic traces.

In Fig.~\ref{fig:logn} we report the average duration of the
observation period required to recover the direct interference graph
for each network size.  The average time is plotted versus the number
of APs for four different traffic intensities $\lambda \in \{0.002,
0.003, 0.004, 0.005\}$. We observe that although, as discussed
  above, the assumptions we adopted to derive the scaling laws do not
  hold in the simulation, the duration required to identify the
  network scales in the predicted, sub-linear (logarithmic), manner in
  network size $n$.  This is consistent with the scaling predicted by
  Thm.~\ref{thm:upper}. The necessary observation time decreases as
  traffic intensity increases, also as predicted by the theory. 

\begin{figure}
\centerline{ \includegraphics[width=9cm]{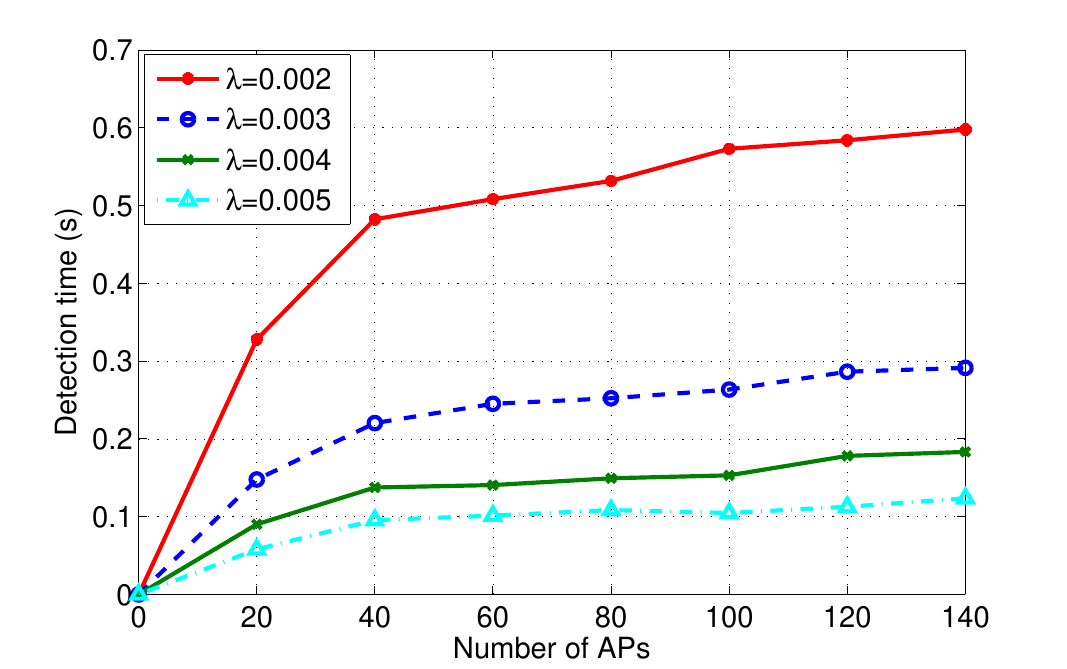}}
\vspace{-0.1in}
\caption{The observation duration required to recover
    the direct interference graph for networks with a maximum of $d=6$
    direct interferers and $s=1$ hidden interferer per node, plotted
    as a function of the number of APs in the network.}
\label{fig:logn}
\end{figure}

\begin{figure}
\centerline{ \includegraphics[width=9cm]{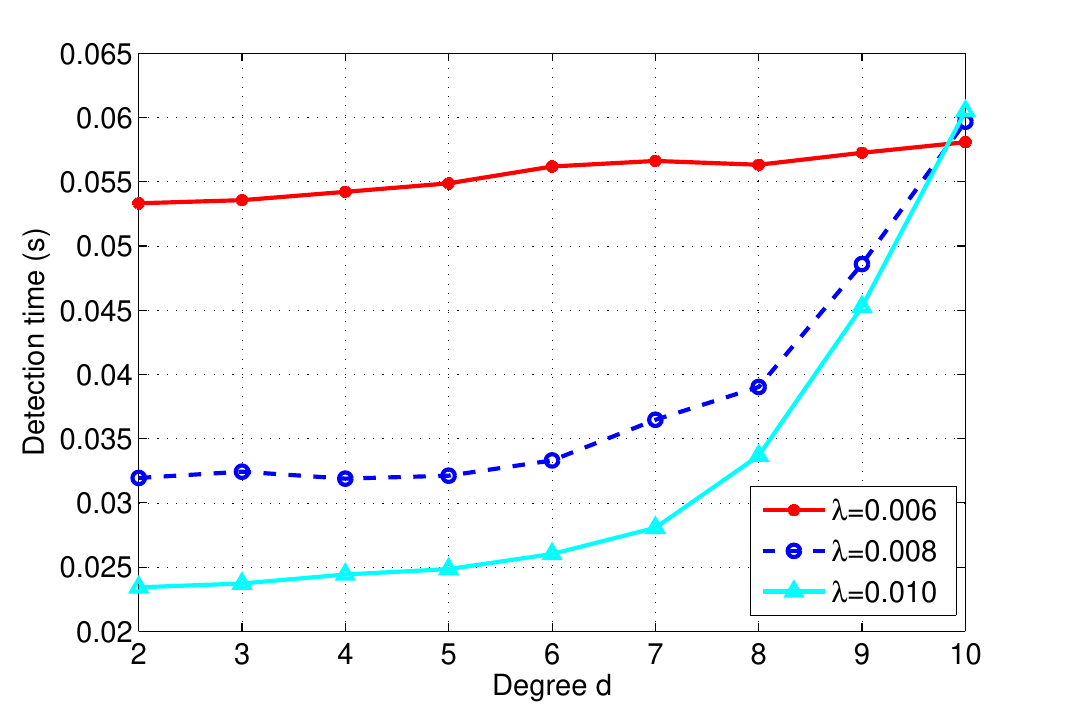}}
 \caption{The observation duration required to recover the direct
   interference graph for networks composed of $4\times 15$ cells,
   plotted as a function of $d$, the maximum number of direct
   interferers.}
\label{fig:cs_d2}
\end{figure}

In Fig.~\ref{fig:cs_d2} we study the average observation time required
to recover the direct interference graph as a function of maximum
degree $d$ for a fixed network size.  We conduct the experiment as
follows.  We fix the network size to be $4\times 15$ cells and
randomly generate topologies (AP and client positions).  For each
randomly generated topology, we vary the carrier sensing range: $25$m, $35$m, $45$m, $55$m, $65$m, $75$m. We check the maximum degree
$d$ (the number of direct interference edges per node) for each
topology. We select ten topologies for each $d$ varying from 2 to
10. We then simulate the network in each case. We plot the average
observation time required to recover $G_D$ as a function of the
maximum degree $d$.

Figure~\ref{fig:cs_d2} demonstrates that in the heavy traffic regime
($\lambda \in \{0.008, 0.01\}$) the scaling is super-linear
(quadratic) in $d$, as predicted by the theory. {However, in a lighter
  traffic regime ($\lambda = 0.006$), the super-linear scaling is not
  obvious.} The reason for the different behavior as a function of
traffic intensity is as follows.  In the heavy traffic regime the
probability that a node competes for the channel does not
increase as $d$ increase since its queue is almost always
  non-empty even when the node does not get to transmit. This makes
the predicted quadratic scaling in $d$ easy to see.  In contrast, in a
sufficiently light traffic regime, the marginal probability that a
node competes for the channel increases as $d$ increases, due to
the time-dependency of the queue state. Since the $p$ in
Thm.~\ref{thm:upper} is now a function of $d$, it essentially
compensates for the quadratic scaling in $d$. Thus, the super-linear
(quadratic) scaling effect is not easily discernible in this regime.

In Fig.~\ref{fig:ack5d2s} we consider the hidden interference graph
estimation problem.  We plot, as a function of network size, the
observation duration required to identify the minimum hitting set
correctly for each node and to recover the hidden interference
graph. The same simulation conditions hold as were described in the
discussion of Fig.~\ref{fig:logn}.  For this algorithm we again
observe that the necessary observation duration scales sub-linearly
(logarithmically) in network size $n$.

In Fig.~\ref{fig:ack5d} we examine the dependence of the necessary
observation duration on $s$, the number of hidden interferers per
node.  In these simulations we fix the network size to be $4\times 15$
cells and the carrier sensing range to be $60$m. We randomly generate
topologies with fixed $d=6$, and let $s$ vary from $1$ to $4$.  The
required observation duration is plotted as a function of $s$.  We see
that the observation duration increases super-linearly as $s$
increases, which is consistent with the predictions of
Thm.~\ref{thm:weak_upper}.

\begin{figure}
\centerline{ \includegraphics[width=9cm]{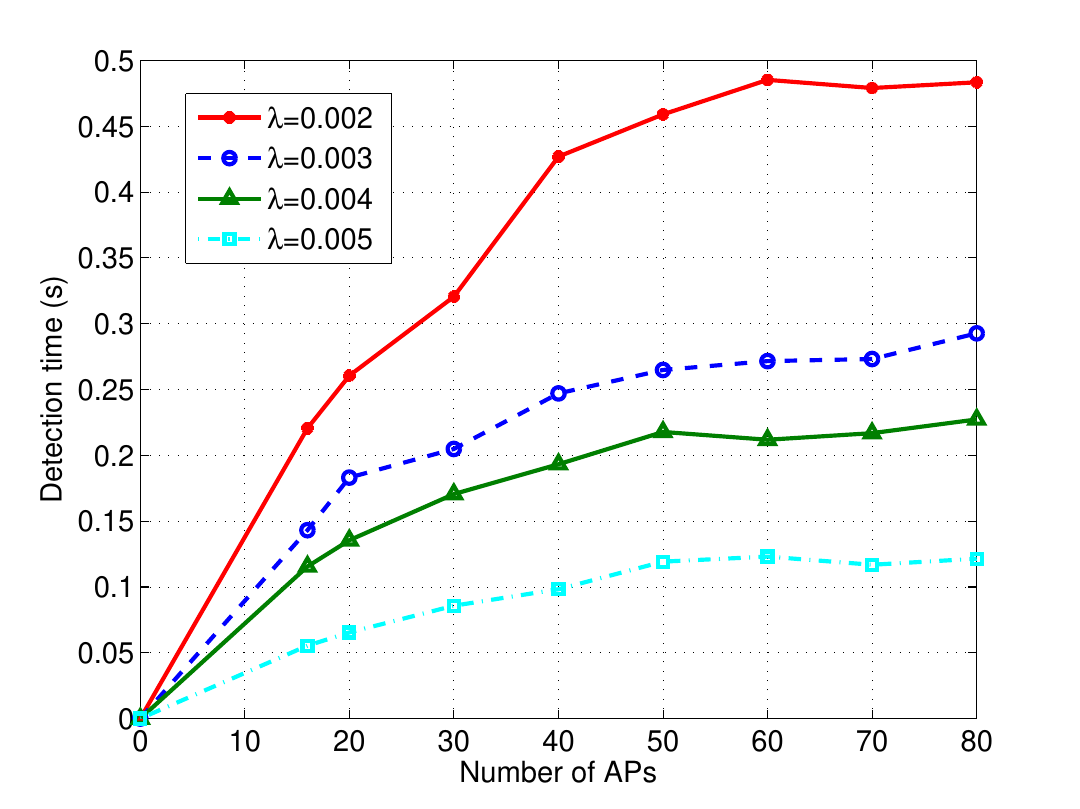}}
\vspace{-0.1in}
\caption{The observation duration required to recover the hidden
  interference graph for networks with a maximum of $d=6$ direct
  interferers and $s=1$ hidden interferer per node, plotted as a function of
  the number of APs in the network.}
\label{fig:ack5d2s}
\end{figure}

\begin{figure}
  \centerline{ \includegraphics[width=9cm]{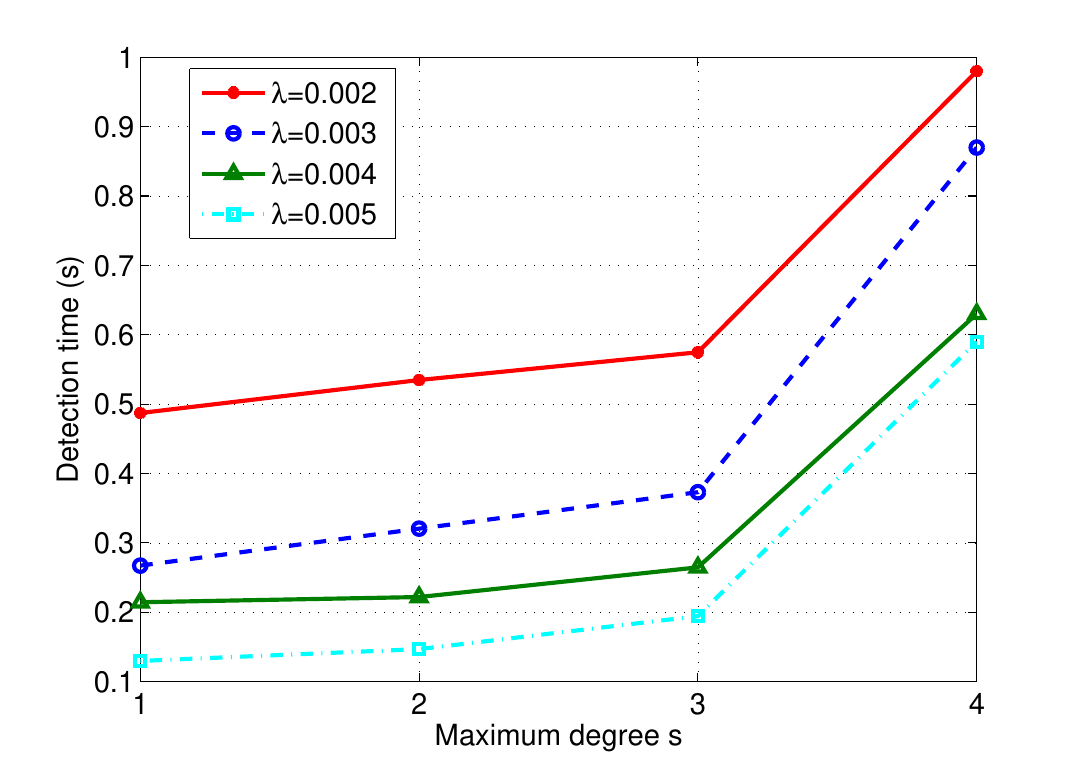}}
 \caption{The observation duration required to recover the hidden
   interference graph for networks composed of $4\times 15$ cells, with a maximum of $d=6$ direct
  interferers per node, plotted as a function of $s$, the maximum number of hidden
   interferers per node.}
\label{fig:ack5d}
\end{figure}

\section{Conclusions}
\label{sec.conclusions}

In this paper, we propose passive interference learning algorithms and
analyze their learning bounds. We first upper bound the number of
measurements required to estimate the direct interference graph. Then,
we provide a minimax lower bound by constructing a sequence of
networks and transforming it into an $M$-ary hypothesis test. The
lower bound matches the upper bound (up to a constant).  Thus, the
bound is tight and the algorithm is asymptotically optimal. We then
analyze the estimation of the hidden interference graph estimation
based on the minimum hitting set algorithm.  We provide matching lower
and upper bounds following an approach similar to that employed for
the direct interference graph.  We also present an experimental study
that lends support to the theoretical analysis.

\appendix
\subsection{Proof of Theorem~\ref{thm:upper}}\label{apx:upper1}
Consider any two nonadjacent nodes $i,j$ in $G_D=(\Vc,\Ec_D)$.  Let
$\mathcal{N}_{ij}=\mathcal{N}_i\cup\mathcal{N}_j$, and
$\mathcal{N}_{i\backslash j}=N_i\cap N_j^c$.

Under the CSMA/CA protocol, transmitter $i$ only contends for the
channel when $Q_i=1$. For ease of exposition, in this proof, we
assume that every transmitter will first choose a random backoff time
at the beginning of each session, whether or not its queue is
non-empty. However, only APs with $Q_i=1$ will actually compete for
the channel.  A transmitter with $Q_i=0$ will not transmit even if it
has the shortest backoff time.

Define $T_{\mathcal{N}_i}$ as the minimum back-off time of the nodes in the set $\mathcal{N}_i$. Then,
\begin{align}
&\Pd\left(T_i<T_{\mathcal{N}_i}, T_j<T_{\mathcal{N}_{j\backslash i}}\right)=\frac{{|\mathcal{N}_{ij}|+2 \choose |\mathcal{N}_i|+1} \cdot |\mathcal{N}_i|! \cdot |\mathcal{N}_{j\backslash i}|!}{(|\mathcal{N}_{ij}|+2)!}\label{eqn:backoff}\\
&\quad=\frac{1}{\left(|\mathcal{N}_i|+1\right)\left(|\mathcal{N}_{ij}|-|\mathcal{N}_{i}|+1\right)}\nonumber\\
&\quad\geq \frac{1}{(d_i+1)(d_j+1)}\geq \frac{1}{(d+1)^2}.\nonumber
\end{align}
The logic underlying~(\ref{eqn:backoff}) is as follows.  Consider
nodes $i,j$ and their neighbors.  There are $|\mathcal{N}_{ij}|+2$
nodes in total, and there are $(|\mathcal{N}_{ij}|+2)!$ orderings of
their back-off times. Among these orderings, ${|\mathcal{N}_{ij}|+2
  \choose |\mathcal{N}_i|+1} \cdot |\mathcal{N}_i|! \cdot
|\mathcal{N}_{j\backslash i}|!$ orderings correspond to
$T_i<T_{\mathcal{N}_i}, T_j<T_{\mathcal{N}_{j\backslash i}}$. Such an
ordering can be obtained in the following way.  Suppose these
$|\mathcal{N}_{ij}|+2$ nodes are ordered according to their back-off
times. Node $i$ and its neighbors take $|\mathcal{N}_i|+1$ positions
in the ordering. There are ${|\mathcal{N}_{ij}|+2 \choose
  |\mathcal{N}_i|+1}$ different combinations of these positions. Since
$T_i<T_{\mathcal{N}_i}$, node $i$ takes the first position out of the
chosen $|\mathcal{N}_i|+1$ positions, the remaining $|\mathcal{N}_i|$
positions are for its neighbors. This results in $|\mathcal{N}_i|!$
orderings for each combinations of positions. Node $j$ and the nodes
in $\mathcal{N}_{j\backslash i}$ take the rest of the positions, where
node $j$ takes the first. This gives the factor
$|\mathcal{N}_{j\backslash i}|!$.

When $T_i<T_{\mathcal{N}_i}$ and $Q_i=1$ then, based on the protocol,
node $i$ gets the channel and thus $X_i=1$. At the same time,
transmissions from all nodes in $\mathcal{N}_i$ are
suppressed. Therefore, for node $j$, if $T_j
<T_{\mathcal{N}_{j\backslash i}}$ and $Q_j=1$, node $j$ also gets a
channel. Thus, we have $X_i=X_j=1$. Since all other scenarios result
in $X_i=X_j=1$, we have the following
\begin{align}
\Pd(&X_i=1,X_j= 1)\nonumber\\
&\geq \Pd\left(T_i<T_{\mathcal{N}_i}, T_j<T_{\mathcal{N}_{j\backslash i}},Q_i=1, Q_j=1\right )\nonumber\\
&= \Pd\left(T_i<T_{\mathcal{N}_i}, T_j<T_{\mathcal{N}_{j\backslash i}}\right) \Pd(Q_i=1, Q_j=1)\nonumber\\
&\geq \frac{p^2}{(d+1)^2}\label{eqn:pair}
\end{align}
and
\begin{align*}
\Pd(&\textrm{edge $(i,j)$ is not removed by a single observation
  $\Xv$})\\ &=1-\Pd(X_i=1,X_j =1)\leq 1-\frac{p^2}{(d+1)^2}.
\end{align*}
Define $\Ac_{ij}$ as the event that edge $(i,j)$ is not removed after $k$ observations.
Then, the probability that, after $k$ observations, the graph cannot be identified successfully is
\begin{align*}
\Pd(\hat{G}_D\neq G_D)&=\Pd( \cup_{(i,j)\notin\Ec_D} \Ac_{ij})\leq {n \choose 2}\left(1-\frac{p^2}{(d+1)^2}\right)^k.
\end{align*}
The inequality follows from the fact that the number of nonadjacent
pairs in $G_D=(\Vc,\Ec_D)$ is upper bounded by ${n \choose 2}$. Under
the assumption that $d\ll n$, this is a good approximation for the
nonadjacent pairs in $G_D$.

\subsection{Proof of Theorem~\ref{thm:low1}}\label{apx1}
Recall that $\mathcal{G}_d$ is defined as the set of graphs consisting
of $n$ nodes that have maximum degree $d$. We are going to construct a
collection of $M+1$ graphs $\{G_{D0}, G_{D1},\ldots, G_{DM}\}$ where
$G_{Di}\in \mathcal{G}_d$ for all $i$. We denote the distribution of
transmission patterns $\xv\in\{0,1\}^n$ for each of these graphs as
$P_0(\xv),P_1(\xv),\ldots,P_M(\xv)$, respectively. Define
$\Pd(\Ac;G_{D})$ to be the probability of event $\Ac$ occurring where
the underlying direct interference graph is $G_D\in \mathcal{G}_d$. A
metric of interest is the edit or ``Levenshtein'' distance between a
pair of graphs.  This is the number of operations (i.e.,
addition/removal of one edge) needed to transform one graph into the
other. We denote the Levenshtein distance between $G_{Di}$ and
$G_{Dj}$ as $D_L(G_{Di},G_{Dj})$. Then, we apply Theorem 2.5 in
  \cite{Tsybakov2008} to obtain the lower bound. We restate the
  theorem in terms of our problem as follows.

\begin{Theorem}\label{thm:minimax} (adapted from~\cite{Tsybakov2008})
  Let $k\in\mathbb{Z}^+$, $M\geq 2$, $\{G_{D0},\ldots,G_{DM}\}\in
  \mathcal{G}_d$ be such that
  \begin{itemize}
  \item[(i)] $D_L(G_{Di},G_{Dj})\geq 2r$, for $0\leq i<j\leq M$, where
    $D_L$ is the
    Levenshtein Distance,
  \item[(ii)]
  $\frac{k}{M}\sum_{i=1}^MD_{KL}(P_i\|P_0)\leq \alpha \log M$, with $0<\alpha<1/8$.
  \end{itemize}Then
  \begin{align*}
    \inf_{\hat{G}_D\in \mathcal{G}_d} & \sup_{G_D\in
      \mathcal{G}_d}\Pd(D_L(\hat{G}_D,G_D)\geq r;G_D)\\ &\geq
    \inf_{\hat{G}_D\in\mathcal{G}_d}\max_{i}\Pd(D_L(\hat{G}_D,G_{Di})\geq
    r;G_{Di})\\ &\geq
    \frac{\sqrt{M}}{1+\sqrt{M}}\left(1-2\alpha-\sqrt{\frac{2\alpha}{\log
        M}}\right)>0.
  \end{align*}
\end{Theorem}

In the following, we apply Theorem~\ref{thm:minimax} to obtain a
  lower bound on $k$. We next construct the $M+1$ graphs $\{G_{D0},
  G_{D1},\ldots, G_{DM}\}$ to satisfy the two conditions -- (i) and
  (ii) -- of Theorem~\ref{thm:minimax}. We first construct $G_{D0}$
and characterize $P_0(\xv)$. Then, we construct the rest of the $M$
graphs by perturbing $G_{D0}$. These graphs will be symmetric in the
sense that $D_{KL}(P_i\|P_0)$ will be the same for $1\leq i\leq M$. We
calculate $D_{KL}(P_1\|P_0)$ and then lower bound the number of
observations $k$ required to determine the interference graph
  with high probability.

\subsubsection{$G_{D0}$ and its transmission pattern distribution $P_0(\xv)$}
Assume $d\geq 2$. We let graph $G_{D0}$ consist of $\lceil n/d\rceil$
disconnected cliques. The first $\lfloor n/d\rfloor:=m_0$ cliques are
each a fully connected subgraph of $d$ nodes, as shown in
Fig.~\ref{fig:int_example1}. The remaining nodes -- if $n/d$ is
  not integer -- form a clique of size less than $d$.  Our analysis
  focuses on the first $m_0$ cliques.

Denote $\Cc_m$ as the set of nodes in the $m$th clique and let
$\Xv_{\Cc_m}$ be the restriction of the transmission pattern $\Xv$ to
the nodes in the clique ($\Xv_{\Cc_m}$ is a set of subvectors, $m =
1,2, \ldots, m_0$ that partition $\Xv$). Define $\ev_i$ as the unit
vector of dimension $d$ whose $i$th element is one. Define $\ov$ to be
the all-zeros vector of length $d$.  Due to the fully connected
structure of the cliques, no more than one node in any $\Cc_m$ can
transmit at any time. We denote the set of all feasible transmission
patterns {for clique $m$ as $\Xc_m:=\{\ov,\ev_1,\ldots,
  \ev_{|\Cc_m|}\}$}. If $\Xv_{\Cc_m}=\ov$, then no node in $\Cc_m$ is
transmitting in that {session}. For each individual clique, this event
happens only when none of the nodes in that clique has traffic to
send, i.e.,
\begin{equation}
\Pd(\Xv_{\Cc_m}=\ov; {G_{D0}})=(1-p)^{d}. \label{eq.noTransmit}
\end{equation}

Otherwise, when at least one AP has a packet to send, the channel will
not idle. Because the $Q_i$s and $T_i$s are i.i.d. across nodes, and
cliques are fully connected, each node in the clique has the same
probability of occupying the channel.  Thus, for any $j$, $j \in \{1,2
\ldots, |\Cc_m|\}$,
\begin{align}
\Pd(\Xv_{\Cc_m}=\ev_j; G_{D0})&=\frac{1-(1-p)^{d}}{d}\triangleq
q, \label{eq.defq}
\end{align}
Since the behavior of the cliques are independent, we have
\begin{align}
P_0(\xv)=\prod_{m=1}^{m_0+1}
\Pd(\Xv_{\Cc_m}=\xv_{\Cc_m};G_{D0}).\label{eq.P0dist}
\end{align}

\subsubsection{Construct $M=n$ graphs}
In this subsection, we construct a sequence of graphs $G_{D1}$,
$G_{D2}$, $\ldots$, $G_{DM}$.  We construct $n$ graphs, i.e., $M = n$.
We construct each graph by picking a pair of nodes from distinct
cliques in first $m_0$ cliques in graph $G_{D0}$.  We add an edge
between the selected pair. We leave the last ($(m_0+1)$th) clique
unmodified for all $G_{D1},\ldots, G_{DM}$. We can construct
${{m_0}\choose 2} d^2 = d^2m_0(m_0-1)/2$ distinct graphs in this
manner. Under the assumption that
\begin{equation}\label{eqn:con1}
d \leq n/2,
\end{equation}
and the fact that $m_0=\lfloor \frac{n}{d}\rfloor\geq \frac{n}{d}-1$,
we lower bound this number of graphs as
\begin{align}
 \frac{d^2}{2}m_0(m_0-1)
 &\geq  \frac{d^2}{2}\left(\frac{n}{d}-1\right)\left(\frac{n}{d}-2\right)\nonumber\\
 &=d^2-\frac{3n}{2}d+\frac{n^2}{2}\label{eqn:num}
\end{align}
 This is a quadratic function of $d$ for
any fixed $n$. We want to construct $M=n$ graphs and the value of (\ref{eqn:num})
equals $n$ if 
\begin{equation}
d= \frac{3n\pm\sqrt{n^2+16n}}{4}. \label{eq.quadEq}
\end{equation}
Since $d\leq n$, and $\sqrt{n^2+16n}>n$, only the smaller solution is
feasible. When $n\geq 7$, we have $\frac{3n-\sqrt{n^2+16n}}{4}>2$,
thus the assumption $d\geq 2$ can be satisfied. Meanwhile, since
\begin{align}
  \frac{3n-\sqrt{n^2+16n}}{4}\leq \frac{3n-n}{4}=\frac{n}{2},
\end{align}
the smaller solution of~(\ref{eq.quadEq}) is a tighter
constraint on $d$ than $(\ref{eqn:con1})$.  Therefore, under the
condition that
\begin{align*}
n\geq 7, \quad 2\leq d\leq \frac{3n-\sqrt{n^2+16n}}{4},
\end{align*}
we have $d^2m_0(m_0-1)/2\geq n$ and thus we can always pick $n$
graphs that are perturbations of $G_{D0}$ in the above sense.  We note
that for each of these graphs $D_L(G_{D0},G_{Di})=1$, and
$D_L(G_{Di},G_{Dj})=2$ for any $0<i,j\leq M$ where $i \neq j$.

\subsubsection{$G_{D1}$ and its transmission pattern distribution $P_1(\xv)$}
We now calculate $P_1(\xv)$, which differs from $P_0(\xv)$ due to the
added constraint resulting from the additional edge. Due to the
symmetric construction of the $M$ graphs, without loss of generality
we concentrate on a single graph.  Let $G_{D1}$ be the graph formed
from $G_{D0}$ by connecting the $i$th node in $\Cc_1$ to the $j$th
node in $\Cc_2$ with an edge.  Since the remaining $m_0-2$ cliques are
unchanged, the distribution of their transmission patterns
$\Xv_{\Cc_m}$ is the same as under $G_{D0}$.  Furthermore, the
remaining transmission patterns are independent of each other and of
$(\Xv_{\Cc_1}, \Xv_{\Cc_2})$.  Thus we express the transmission
pattern distribution under $G_{D1}$ as
\begin{align}
P_1(\xv)&=\Pd(\Xv_{\Cc_1}=\xv_{\Cc_1},\Xv_{\Cc_2}=\xv_{\Cc_2};G_{D1})\nonumber\\
&\quad\cdot\prod_{m=3}^{m_0+1}\Pd(\Xv_{\Cc_m}=\xv_{\Cc_m};G_{D0}). \label{eq.P1dist}
\end{align}
We want to calculate the KL-divergence between $P_0(\xv)$
from~(\ref{eq.P0dist}) and $P_1(\xv)$.  The final $m_0-2$ terms of
both are identical.  Thus, for the remainder of this subsection we
focus on the distribution of the activation pattern in the first two
cliques, i.e., $ \Pd(\Xv_{\Cc_1}=\xv_{\Cc_1},
\Xv_{\Cc_2}=\xv_{\Cc_2};G_{D1})$.

\begin{figure}
\centerline{ \includegraphics[width=7cm]{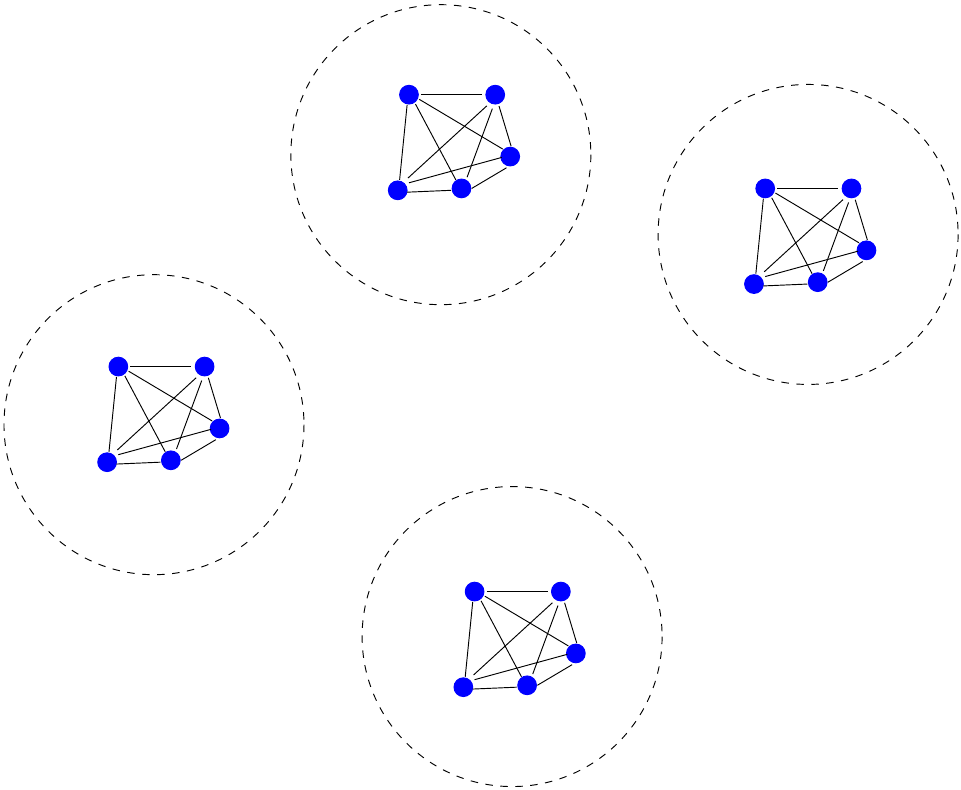}}
\caption{A network consists of $n/d$ cliques, where nodes in each
  clique are within each other's carrier sensing range. Nodes
  belonging to different cliques cannot hear each other. $n=20,d=5$.}
\label{fig:int_example1}
\vspace*{-0.15in}
\end{figure}

The differences between $ \Pd(\Xv_{\Cc_1}=\xv_{\Cc_1},
\Xv_{\Cc_2}=\xv_{\Cc_2};G_{D1})$ and $ \Pd(\Xv_{\Cc_1}=\xv_{\Cc_1},
\Xv_{\Cc_2}=\xv_{\Cc_2};G_{D0})$ are due to the added edge, which
constraints the allowable patterns.  In particular, the event
$\Xv_{\Cc_1}=\ev_i, \Xv_{\Cc_2}=\ev_j$, which occurs with nonzero
probability under $G_{D0}$, cannot occur under $G_{D1}$. If, under
$G_{D0}$ both nodes $i$ and $j$ would have transmitted, under $G_{D1}$
only one will transmit.  The carrier sensing mechanism will suppress
the transmission of the other.  Which node will transmit and which
will be suppressed depends on the respective back-off times. The
result will be an increase in the probability of some other node
transmitting, since the suppressed node will not compete for the
channel.

To get the analysis rolling, we first consider the simpler situations
of joint transmission patterns $(\xv_{\Cc_1}, \xv_{\Cc_2})$ where
there is no possibility that transmission by node $i$ could have
suppressed transmission by node $j$ (or vice-versa).  For these
situations the probability of the joint pattern under $G_{D1}$ is the
same as under $G_{D0}$.  There are three such cases.  The first case
consists of patterns such that (i) some node in $\Cc_i$ {\em other
  than} $i$ transmits and (ii) no node in $\Cc_2$ transmits, i.e.,
$\xv_{\Cc_1}\in\Xc_1\backslash\ev_i$ and $\xv_{\Cc_2}=\ov$.  Condition
(i) means that node $i$ could not have suppressed the transmission of
node $j$.  Thus, by condition (ii) no node in $\Cc_2$ has any data to
transmit.  Therefore,
\begin{equation*}
\Pd(\Xv_{\Cc_1} \tEq \xv_{\Cc_1}, \Xv_{\Cc_2}\tEq\ov;G_{D1}) \tEq
\Pd(\Xv_{\Cc_1} \tEq \xv_{\Cc_1}, \Xv_{\Cc_2}\tEq\ov;G_{D0}).
\end{equation*}
The second case is the reverse of the first, i.e., (i)
$\xv_{\Cc_2}\in\Xc_2\backslash\ev_j$ and (ii) $\xv_{\Cc_1}=\ov$.  By
the same logic, $\Pd(\Xv_{\Cc_1}=\ov, \Xv_{\Cc_2}=\xv_{\Cc_2};G_{D1})
=\Pd(\Xv_{\Cc_1}=\ov, \Xv_{\Cc_2}=\xv_{\Cc_2};G_{D0})$.  The third
case consists of situations where (i) transmissions occur in both
cliques, but (ii) neither $i$ nor $j$ transmit.  Condition (ii)
implies that suppression of node $j$ by node $i$ (or $i$ by $j$) could
not have occurred, and thus the probability of the joint pattern under
$G_{D1}$ is the same as under $G_{D0}$.  These conditions are
summarized as $\xv_{\Cc_1}\in \Xc_1 \backslash \{\ov,\ev_i\}$ and
$\xv_{\Cc_2}\in \Xc_2\backslash \{\ov,\ev_j\}$.  Thus,
 \begin{equation*}
 \Pd(\Xv_{\Cc_1}\!\tEq\xv_{\Cc_1},
 \Xv_{\Cc_2}\!\tEq\xv_{\Cc_2};G_{D1})
 \tEq\Pd(\Xv_{\Cc_1}\!\tEq\xv_{\Cc_1},
 \Xv_{\Cc_2}\!\tEq\xv_{\Cc_2};G_{D0}).
 \end{equation*}
Finally, by~(\ref{eq.defq}) and~(\ref{eq.P0dist}) we know that
$\Pd(\Xv_{\Cc_1}\!\tEq\xv_{\Cc_1},
\Xv_{\Cc_2}\!\tEq\xv_{\Cc_2};G_{D0}) = q^2$.

We now turn to transmission patterns where nodes $i$ and $j$ interact.
These are patterns $(\xv_{\Cc_1}, \xv_{\Cc_2})$ the probability of
which may be higher under $G_{D1}$ than under $G_{D0}$ due the the
infeasibility of the pattern $(\ev_i, \ev_j)$ under $G_{D1}$.  To
analyze these patterns consider the paired queue/backoff-time vectors
$(\Qv_{ \Cc_1\cup\Cc_2}, \Tv_{\Cc_1\cup\Cc_2})$ that, under $G_{D0}$,
would have resulted in $(\Xv_{\Cc_1}, \Xv_{\Cc_2}) = (\ev_i,\ev_j)$. Obviously,
we must have $\mathbf{Q}_{\Cc_1},\mathbf{Q}_{\Cc_2}\neq \bf{0}$.
Depending on the particular realization of $(\Qv_{ \Cc_1\cup\Cc_2},
\Tv_{\Cc_1\cup\Cc_2})$ there are four possible outcomes under
$G_{D1}$:
\begin{itemize}
\item[(a)] $T_i<T_j$ and $\mathbf{Q}_{\Cc_2}=\ev_j$: Node $i$ will get
  the channel and transmit so $\Xv_{\Cc_1}=\ev_i$. However, the
  transmission of node $j$ will be suppressed. Since
  $\mathbf{Q}_{\Cc_2}=\ev_j$, node $j$ is the only node in $\Cc_2$
  with data, so $\Xv_{\Cc_2}=\ov$.
\item[(b)] $T_i>T_j$ and $\mathbf{Q}_{\Cc_1}=\ev_i$: The analysis here
  is analogous to (a) with the roles of $i$ and $j$ reversed.  Thus,
  the transmission pattern will be
  $\Xv_{\Cc_1}=\ov,\Xv_{\Cc_2}=\ev_j$.
\item[(c)] $T_i<T_j$ and $\mathbf{Q}_{\Cc_2}\neq\ev_j$: In this
  situation at least one other node in $\Cc_2$ has data to transmit.
  Thus, even though $\Xv_{\Cc_1}=\ev_i$ suppresses the transmission of
  node $j$, some other node in $\Cc_2$ will transmit.  Thus,
  $\Xv_{\Cc_2} \in \Xc_2\backslash\{\ev_j,\ov\}$.  Furthermore, since
  the statistics of the queue statuses and backoff times are identical
  for all nodes, the transmitting node in $\Cc_2$ will be uniformly
  distributed across the other $d-2$ nodes in that clique.
\item[(d)] $T_i>T_j$ and $\mathbf{Q}_{\Cc_1}\neq\ev_i$: The analysis
  here is analogous to (c) with the roles of $i$ and $j$ reversed.
  $\Xv_{\Cc_2}=\ev_j$, and $\Xv_{\Cc_1}$ is uniformly distributed
  across $\xv_{\Cc_1}\in \Xc_1\backslash\{\ev_i,\ov\}$.
\end{itemize}
We note that the above four cases
partition the event space where nodes $i$ and $j$ transmit
concurrently under $G_{D0}$.  The four terms in the following
correspond, respectively, to (a)--(d), above:
\begin{align}
\Pd&(\Xv_{\Cc_1}=\ev_i,
\Xv_{\Cc_2}=\ev_j;G_{D0})\nonumber\\
=& \ \Pd(\Xv_{\Cc_1}=\ev_i, \Xv_{\Cc_2}=\ev_j,T_i<T_j,
\mathbf{Q}_{\Cc_2}=\ev_j;G_{D0})\nonumber\\
&+\Pd(\Xv_{\Cc_1}=\ev_i, \Xv_{\Cc_2}=\ev_j,T_i>T_j,
\mathbf{Q}_{\Cc_1}=\ev_i;G_{D0})\nonumber\\
&+\Pd(\Xv_{\Cc_1}=\ev_i, \Xv_{\Cc_2}=\ev_j,T_i<T_j,
\mathbf{Q}_{\Cc_2}\neq\ev_j;G_{D0})\nonumber\\
&+\Pd(\Xv_{\Cc_1}=\ev_i, \Xv_{\Cc_2}=\ev_j,T_i>T_j,
\mathbf{Q}_{\Cc_1}\neq\ev_i;G_{D0})\nonumber\\
=& \ 2 \beta_1 + 2 \beta_2.\label{eqn:p}
\end{align}
The first two terms are equal due to the symmetry of the conditions
and graph structure.  Similar logic implies that the third and fourth
terms are also equal.  We respectively define $\beta_1$ and $\beta_2$
to be the two probabilities.

We now consider the four cases of joint transmission patterns
$(\xv_{\Cc_1}, \xv_{\Cc_2})$ under $G_{D1}$ not yet considered.  These
will each connect to one of the cases (a)--(d), above.  First,
consider the probability of pattern $(\xv_{\Cc_1}, \xv_{\Cc_2})
=(\ev_i,\ov)$ under $G_{D1}$.  The probability of this pattern under
$G_{D1}$ will be larger than under $G_{D0}$ since certain pairs
$(\Qv,\Tv)$ that result in $(\ev_i, \ev_j)$ under $G_{D0}$ result in
to $(\ev_i, \ov)$ under $G_{D1}$.  The probability of observing
$(\ev_i, \ov)$ under $G_{D1}$ equals the probability of observing that
pattern under $G_{D0}$ plus the probability of the event occurring
that was considered in case (a), above.  This latter event is the bump
in probability due to the interaction of $i$ and $j$.  Therefore, we
find that
\begin{align}
&\Pd(\Xv_{\Cc_1}=\ev_i,
  \Xv_{\Cc_2}=\ov;G_{D1})\nonumber\\ &=\Pd(\Xv_{\Cc_1}=\ev_i,
  \Xv_{\Cc_2}=\ov;G_{D0})\nonumber\\ &\quad+\Pd(\Xv_{\Cc_1}=\ev_i,
  \Xv_{\Cc_2}=\ev_j,T_i<T_j,
  \mathbf{Q}_{\Cc_2}=\ev_j;G_{D0})\nonumber\\ &=q(1-p)^{d}+\beta_1, \label{eq.useBeta1_a}
\end{align}
where the first term follows from the independence of the node-wise
transmission patterns under $G_{D0}$, from the definition of the
  probability $q$ in~(\ref{eq.defq}), and from the fact that with
  probability $(1-p)^d$ no nodes in $\Cc_2$ have data to transmit.
We defer the calculation of $\beta_1 = \Pd(\Xv_{\Cc_1}=\ev_i,
\Xv_{\Cc_2}=\ev_j,T_i<T_j, \mathbf{Q}_{\Cc_2}=\ev_j;G_{D0})$
until~(\ref{eqn:backoff2}). Following a similar line of argument for
$(\xv_{\Cc_1}, \xv_{\Cc_2}) =(\ov, \ev_j)$, and considering case (b),
we find that
\begin{equation}
\Pd(\Xv_{\Cc_1}=\ov, \Xv_{\Cc_2}=\ev_j;G_{D1})=q(1-p)^{d}+\beta_1. \label{eq.useBeta1_b}
\end{equation}

The third case concerns the patterns $(\ev_i, \xv_{\Cc_2})$ for all
$\xv_{\Cc_2}\in\Xc_2\backslash\{\ov,\ev_j\}$.  Similar to (a) and (b),
the probability of a pattern in this set will be at least as
  large as the probability of the pattern under $G_{D0}$ due to a
  boost in probability resulting from the infeasibility of the
$(\ev_i, \ev_j)$ pattern under $G_{D1}$.  The boost corresponds to the
event discussed in (c), above.  For any $\xv_{\Cc_2}\in
\Xc_2\backslash \{\ev_j,\ov\}$ we find that
\begin{align}
&\Pd(\Xv_{\Cc_1}=\ev_i,\Xv_{\Cc_2}=\xv_{\Cc_2};G_{D1})\nonumber\\ &=\Pd(\Xv_{\Cc_1}=\ev_i,\Xv_{\Cc_2}=\xv_{\Cc_2};G_{D0})\nonumber\\ &\quad+\frac{1}{d-2}\Pd(\Xv_{\Cc_1}=\ev_i,
  \Xv_{\Cc_2}=\ev_j,T_i<T_j, \mathbf{Q}_{\Cc_2}\neq
  \ev_j;G_{D0})\nonumber\\ &=q^2+{\frac{1}{d-1}}\beta_2. \label{eq.useBeta2_a}
\end{align}
The factor of $1/(d-1)$ in the second term results from the uniformity
over the other $d-1$ transmission patterns in $\Cc_2$, mentioned in
(c).  The probability $\beta_2 =\Pd(\Xv_{\Cc_1}=\ev_i,
\Xv_{\Cc_2}=\ev_j,T_i<T_j, \mathbf{Q}_{\Cc_2}\neq\ev_j;G_{D0})$ will
be calculated in~(\ref{eq.defBeta2}). Finally, by symmetric logic, we
find that for any $\xv_{\Cc_1}\in \Xc_1\backslash \{\ev_i,\ov\}$
\begin{equation}
\Pd(\Xv_{\Cc_1}=\xv_{\Cc_1},\Xv_{\Cc_2}=\ev_j;G_{D1})=q^2+{\frac{1}{d-1}}\beta_2. \label{eq.useBeta2_b}
\end{equation}

We now calculate $\beta_1$, required in~(\ref{eq.useBeta1_a})
and~(\ref{eq.useBeta1_b}).  We start by rewriting the first
term~(\ref{eqn:p}) using Bayes' rule as
\begin{align*}
\Pd(&\Xv_{\Cc_1}\!=\!\ev_i, \Xv_{\Cc_2}\!=\!\ev_j,T_i<T_j,
\mathbf{Q}_{\Cc_2}\!=\!\ev_j;G_{D0}) \nonumber\\ &= \Pd(\Xv_{\Cc_1}\!=\!\ev_i,
T_i<T_j;G_{D0}) \Pd(\Xv_{\Cc_2}\!=\!\ev_j,
\mathbf{Q}_{\Cc_2}\!=\!\ev_j;G_{D0}).
\end{align*}
In the application of Bayes' rule we have used the fact that
$\Pd(\Xv_{\Cc_1}=\ev_i, T_i<T_j | \Xv_{\Cc_2}=\ev_j,
\mathbf{Q}_{\Cc_2}=\ev_j;G_{D0}) = \Pd(\Xv_{\Cc_1}=\ev_i,
T_i<T_j;G_{D0})$, due to the independence of transmission patterns
under $G_{D0}$.  Now, note that $\Pd(\Xv_{\Cc_2}=\ev_j,
\mathbf{Q}_{\Cc_2}=\ev_j;G_{D0}) = p (1-p)^{d-1}$ as node $j$ is the
only node in $\Cc_2$ with something to transmit. Define $\Tc_{1,i}=\{j
\in \Cc_1 : T_j < T_i\}$, i.e., the set of APs in $\Cc_1$ whose
backoff time is shorter than that of AP $i$.  Next, rewrite
$\Pd(\Xv_{\Cc_1}=\ev_i, T_i<T_j;G_{D0})$ as
\begin{align*}
&\sum_{l=0}^{d-1} \Pd(\Xv_{\Cc_1}=\ev_i, T_i<T_j, |\Tc_{1,i}| = l;G_{D0})\\
& = \sum_{l=0}^{d-1} \Pd\Big(\Xv_{\Cc_1}=\ev_i \Big| T_i<T_j, |\Tc_{1,i}| = l;G_{D0}\Big)\\
& \hspace{1em} \cdot \Pd(T_i<T_j, |\Tc_{1,i}| =
  l;G_{D0}).
\end{align*}
The first factor $\Pd\Big(\Xv_{\Cc_1}=\ev_i \Big| T_i<T_j, |\Tc_{1,i}|
= l;G_{D0}\Big) = p (1-p)^l$.  The second factor is just the fraction
of the $d !$ orderings such that there are $l$ nodes in $\Cc_1$ with
backoff times lower than $T_i$ and such that node $T_i < T_j$.  The
number of such orderings is ${{d-1}\choose l} l! (d-l)!  = (d-1)!
(d-l)$.  Putting the pieces together we find that
\begin{align}
\beta_1&\triangleq  \Pd(\Xv_{\Cc_1}=\ev_i, \Xv_{\Cc_2}=\ev_j,T_i<T_j,  \mathbf{Q}_{\Cc_2}=\ev_j;G_{D0})\nonumber\\
&= \Pd(\Xv_{\Cc_1}\!=\!\ev_i, T_i<T_j;G_{D0})\Pd(\Xv_{\Cc_2}\!=\!\ev_j, \mathbf{Q}_{\Cc_2}\!=\!\ev_j;G_{D0})\nonumber\\
  &=\left[\sum_{l=0}^{d-1}(d-l)(1-p)^l\right]
\frac{p^2(1-p)^{(d-1)}}{d(d+1)} \\
&=\frac{\left(p(d+1)-1+(1-p)^{d+1}\right)(1-p)^{(d-1)}}{d(d+1)}.
\label{eqn:backoff2}
\end{align}
And, since transmitters $i$ and $j$ have the same statistics, when
$T_i>T_j$, we also have
\begin{align*}
 & \Pd(\Xv_{\Cc_1}=\ev_i,\Xv_{\Cc_2}=\ev_j,T_i>T_j,  \mathbf{Q}_{\Cc_1}=\ev_i;G_{D0})=\beta_1,
\end{align*}
which justifies~(\ref{eq.useBeta1_b}).

Finally, to calculate $\beta_2$ we simply combine~(\ref{eqn:p})
with~(\ref{eqn:backoff2}).
\begin{align*}
  \Pd(\Xv_{\Cc_1}=\ev_i, \Xv_{\Cc_2}=\ev_j;G_{D0})&=2\beta_1+2\beta_2=q^2,
\end{align*}
where the final inequality follows from the independence of
$\Xv_{\Cc_1}$ and $\Xv_{\Cc_2}$ under $G_{D0}$.  Thus,
\begin{equation}
\beta_2=\frac{q^2-2\beta_1}{2}. \label{eq.defBeta2}
\end{equation}

\subsubsection{Bounding $D_{KL}(P_0\|P_1)$}
The Kullback-Leibler divergence between the distribution of
transmission patterns under $G_{D1}$ and $G_{D0}$, denoted as $P_1$
and $P_0$, respectively, can be calculated as
\begin{align}
 & D_{KL}(P_1\|P_0) = \sum_{\xv\in
    \{0,1\}^n}P_1(\xv)\log \frac{P_1(\xv)}{P_0(\xv)}\nonumber\\
 &= \ \sum_{\xv_{\Cc_1} \in \{0,1\}^{d}, \xv_{\Cc_2} \in
    \{0,1\}^{d}} \hspace{-3em}
  \Pd(\Xv_{\Cc_1}=\xv_{\Cc_1},\Xv_{\Cc_2}=\xv_{\Cc_2};G_{D1})
  \nonumber \\ & \hspace{2em} \cdot \log
  \frac{\Pd(\Xv_{\Cc_1}=\xv_{\Cc_1},\Xv_{\Cc_2}=\xv_{\Cc_2};G_{D1})}{\Pd(\Xv_{\Cc_1}
    = \xv_{\Cc_1};G_{D0})
    \Pd(\Xv_{\Cc_2}=\xv_{\Cc_2};G_{D0})} \label{eq.cancelCommonTerms}\\
 &= \ 2 \Pd(\Xv_{\Cc_1}=\ev_i,\Xv_{\Cc_2}=\ov;G_{D1})
  \nonumber \\ & \hspace{2em} \cdot \log
  \frac{\Pd(\Xv_{\Cc_1}=\ev_i,\Xv_{\Cc_2}=\ov;G_{D1})}{\Pd(\Xv_{\Cc_1}=\ev_i;G_{D0})
    \Pd(\Xv_{\Cc_2}=\ov;G_{D0})}\nonumber\\
& \quad+ 2 (d-1) \Pd(\Xv_{\Cc_1}=\ev_i,\Xv_{\Cc_2}=\ev_{j'};G_{D1})
  \nonumber \\ & \hspace{2em} \cdot \log
  \frac{\Pd(\Xv_{\Cc_1}=\ev_i,\Xv_{\Cc_2}=\ev_{j'};G_{D1})}{\Pd(\Xv_{\Cc_1}=\ev_i;G_{D0})
    \Pd(\Xv_{\Cc_2}=\ev_{j'};G_{D0})} \label{eq.imptTerms}    \\
 &= \ 2[q(1-p)^{d}+\beta_1]\log\left[\frac{q(1-p)^{d}+\beta_1}{q(1-p)^{d}}\right]
\nonumber\\ &\quad +
2(d-1)\left[q^2+\frac{\beta_2}{d-1}\right]\log\left[\frac{q^2+\frac{\beta_2}{d-1}}{q^2}\right]\label{eq.evaluate}\\
 &\leq \frac{2\beta_1}{q(1-p)^{d}}(q(1-p)^{d}\hspace{-0.03in}+\hspace{-0.03in}\beta_1)+ \frac{2\beta_2}{q^2}\left(q^2\hspace{-0.03in}+\hspace{-0.03in}\frac{\beta_2}{d\hspace{-0.03in}-\hspace{-0.03in}1}\right)\label{eqn:log}\\
 &=q^2+\frac{2\beta_1^2}{q(1-p)^{d}}+\frac{2\beta_2^2}{(d-1)q^2}.\label{eqn:2b0}
 \end{align}
In~(\ref{eq.cancelCommonTerms}) we cancel (and marginalize over) the
$m_0 - 1$ common factors of the form
$\Pd(\Xv_{\Cc_m}=\xv_{\Cc_m};G_{D0})$, $3 \leq m \leq m_0+1$
cf.~(\ref{eq.P0dist}) and~(\ref{eq.P1dist}).  In~(\ref{eq.imptTerms}),
we focus on the terms that don't cancel out.  The first two terms
therein correspond to cases (a) and (b), cf.~(\ref{eq.useBeta1_a})
and~(\ref{eq.useBeta1_b}).  The latter $2(d-2)$ terms correspond the
cases (c) and (d), cf.~(\ref{eq.useBeta2_a})
and~(\ref{eq.useBeta2_b}), where $j'$ is some node $j' \in \Cc_2$ but
$j' \neq j$.  In~(\ref{eq.evaluate}) we use~(\ref{eq.useBeta1_a})
and~(\ref{eq.useBeta2_a}) in the numerators and~(\ref{eq.noTransmit})
and~(\ref{eq.P0dist}) in the denominators. The inequality
in~(\ref{eqn:log}) follows from the fact that $\log (1+x)\leq
x$. Based on the definitions of $q$, $\beta_1$ and $\beta_2$
from~(\ref{eq.defq}),~(\ref{eqn:backoff2}) and~(\ref{eq.defBeta2}), we
have
\begin{align}
&\frac{2\beta_1^2}{q(1-p)^{d}}=\frac{2\beta_1}{q}\frac{\beta_1}{(1-p)^{d}}\nonumber\\
&=\frac{2\beta_1}{1-(1-p)^{d}}\frac{1}{d+1}\left(\frac{dp}{1-p}-1+(1-p)^{d}\right)\nonumber\\
&\leq \frac{2\beta_1}{d+1}\left(\frac{d}{1-(1-p)^1}\cdot\frac{p}{1-p}-1\right)\leq \frac{2\beta_1}{1-p}\label{eqn:2b1}
\end{align}
and
\begin{align}
&\frac{2\beta_2^2}{(d-1)q^2}=\frac{\beta_2}{d-1}\frac{2\beta_2}{q^2}\leq \beta_2\leq q^2.\label{eqn:2b}
\end{align}
Plugging (\ref{eqn:2b1}) and (\ref{eqn:2b}) into (\ref{eqn:2b0}), we have
    \begin{align}
    D_{KL}(P_1\|P_0)
 \leq & \left(2+\frac{1}{1-p}\right)\frac{1}{d^2}.\label{eqn:2b2}.
\end{align}

\subsubsection{Put pieces together}
Since the $G_{Di}$s are constructed in the same manner,
$D_{KL}(P_i\|P_0) = D_{KL}(P_1\|P_0)$ for all $i$.  Thus, we have
\begin{align*}
  \frac{1}{M}\sum_{i=1}^M D_{KL}(P_i\|P_0)&\leq
  \left(2+\frac{1}{1-p}\right)\frac{1}{d^2}.
\end{align*}

In summary, we have $D_L(G_{Di},G_{Dj})\geq 1$ for $0\leq i<j\leq M$
and we can always pick $M=n$. Thus, according to
Thm.~\ref{thm:minimax}, when
\begin{align*}k&\leq  \frac{ \alpha\log n}{\left(2+\frac{1}{1-p}\right)\frac{1}{d^2}}
\end{align*} we have
\begin{align*}
 \inf_{{\hat{G}_D\in \mathcal{G}_d}}&\sup_{G_D\in \mathcal{G}_d}\Pd(\hat{G}_D\neq G_D;G_D )\\
 &= \inf_{{\hat{G}_D\in \mathcal{G}_d}}\sup_{G_D\in \mathcal{G}_d}\Pd(D_L(\hat{G}_D,G_D)\geq 1/2;G_D)\\
 &>\frac{\sqrt{n}}{1+\sqrt{n}}\left(1-2\alpha-\sqrt{\frac{2\alpha}{\log n}}\right)>0.
\end{align*}

\subsection{Proof of Theorem~\ref{thm:weak_upper}}\label{apx:weak_upper}
Before we prove Theorem~\ref{thm:weak_upper}, we first show that the
following lemma is true.
\begin{Lemma}\label{lemma:minhittingset}
$\lim_{k\rightarrow \infty}\Pd(\hat{\Sc}_j(k)= \Sc_j)=1$.
\end{Lemma}

The lemma states that when $k$ is sufficiently large, identifying the
minimum hitting set of the candidate interferer sets is equivalent to
identifying the hidden interferer set of an AP.

\begin{Proof}
Since $Y_j=0$ must be caused by some active interferer, $\Sc^t_j\cap
\Sc_j\neq \varnothing$ for every $t\in \Kc_j(k)$. Therefore, $\Sc_j$
is a hitting set for $\{\Sc_j^t\}_{t\in\Kc_j(k)}$.

Next, we prove that $\Sc_j$ is the unique minimum hitting set. We
prove this through contradiction. Assume that there exist a different
hitting set $\hat{\Sc}_j(k)$ with $|\hat{\Sc}_j(k)|\leq
|\Sc_j|$. Since $\hat{\Sc}_j(k)$ is different from $\Sc_j$, there must
exist a node $i\in\Sc_j$, that is not in $\hat{\Sc}_j(k)$, i.e., $i\in
\Sc_j\backslash \hat{\Sc}_j(k)$. Consider the following probability
\begin{align}
&\Pd(X_i=1,X_j=1,Y_j=0,\Xv_{\hat{\Sc}_j(k)}=\ov)\nonumber\\ &=\Pd(X_i=1,X_j=1,\Xv_{\hat{\Sc}_j(k)}=\ov)\nonumber\\ &\qquad\cdot\Pd(Y_j=0|X_i=1,X_j=1,
  \Xv_{\hat{\Sc}_j(k)}=\ov)\nonumber\\ &\geq
  \Pd(X_i=1,X_j=1,\Qv_{\hat{\Sc}_j(k)}=\ov)\nonumber\\ &\qquad\cdot\Pd(Y_j=0|X_i=1,X_j=1,
  \Xv_{\hat{\Sc}_j(k)}=\ov)\nonumber\\
&=\Pd(X_i=1,X_j=1|\Qv_{\hat{\Sc}_j(k)}=\ov)\cdot\Pb(\Qv_{\hat{\Sc}_j(k)}=\ov) \nonumber\\ &\qquad\cdot\Pd(Y_j=0|X_i=1,X_j=1,
  \Xv_{\hat{\Sc}_j(k)}=\ov)\nonumber\\
   &\geq \frac{p^2}{(d+1)^2}(1-p)^{|\hat{\Sc}_j(k)|}p_{ij} 
   \geq \frac{p^2}{(d+1)^2}(1-p)^{s} p_{\min},\label{eqn:lemma12}
\end{align}
where the first inequality in~(\ref{eqn:lemma12}) is based on the
  observation that given $\Qv_{\hat{\Sc}_j(k)}$, the network behaves
  as if the APs in $\hat{\Sc}_j(k)$ do not exist at all. With fewer
  nodes and possibly fewer edges in the corresponding direct
  interference graph, the bound we derive in (\ref{eqn:pair}) is still
  valid. The second inequality in~(\ref{eqn:lemma12}) follows from the assumptions
that $i\in \Sc_j$, $|\hat{\Sc}_j(k)|\leq |\Sc_j|\leq s$, and
from~(\ref{eqn:pij_bound}).

For any observation with
  $X_i(t)=1,X_j(t)=1,Y_j(t)=0,\Xv_{\hat{\Sc}_j(k)}(t)=\ov$, the time
  index $t$ is an index of $\Kc_j(k)$. However, $\Sc^t_j\cap
  \hat{\Sc}_j(k)=\varnothing$.  This contradicts the assumption that
  $\hat{\Sc}_j(k)$ is a hitting set. Since the probability of this
  event has a lower bound for any fixed $d$ and $s$, this event
  happens with probability one as $k\rightarrow\infty$. Therefore,
  $\Sc_j$ is the unique minimum hitting set as $k\rightarrow\infty$.
\end{Proof}

Define the error event $E_j$ as the event that the estimated minimum
hitting set $\hat{\Sc}_j(k)$ is not equal to $\Sc_j$ after $k$
observations. This only happens when $|\hat{\Sc}_j(k)|\leq
|{\Sc}_j|$. So when $|\hat{\Sc}_j(k)|\leq |{\Sc}_j|$, there must exist
at least one transmitter $i\in \Sc_j$ that is not included in
$\hat{\Sc}_j(k)$. Then, following steps similar to those followed in
the proof of Lemma~\ref{lemma:minhittingset}, we have
\begin{align*}
\Pd(E_j)&=\Pd(\cup_{i\in \Sc_j} i\notin \hat{\Sc}_j(k))\\
&\leq\sum_{i\in {\Sc}_j }(1-\Pd(X_i=1,X_j=1,Y_j=0,\Xv_{\hat{\Sc}_j(k)}=\ov))^k\\
&\leq s\left(1-\frac{p^2}{(d+1)^2}(1-p)^{s}p_{\min}\right)^k.
\end{align*}
Therefore,
 \begin{align*}
\Pd(\hat{G}_H\neq G_H) = \Pd(\cup _{j}E_{j}) & \leq \sum_{j}s\left(1\!-\!\frac{p^2}{d^2}(1\!-\!p)^{s-1}p_{\min}\right)^k\\
  &= ns\left(1-\frac{p^2(1-p)^{s}p_{\min}}{(d+1)^2}\right)^k.
\end{align*}

\subsection{Proof of Theorem~\ref{thm:low3}}\label{apx:low3}

To prove Theorem~\ref{thm:low3} we follow a similar approach to that
taken in the proof of Theorem~\ref{thm:low1}. For any given direct
interference graph $G_D=(\Vc,\Ec_D)$, we define $\mathcal{H}_s(G_D)$
to be the set of hidden interference graphs satisfying the assumption
that $s_j\leq s$ for every $j$.  We construct a collections of graphs,
$G_{H0}, G_{H1},\ldots, G_{HM}$, all in $\mathcal{H}_s(G_D)$, and
reduce the interference graph estimation problem to an $M$-ary
hypothesis test. These graphs share the same node set and direct
interference edges, however, the hidden interference edges
differ. With slight abuse of the notation, we use $P_i(\xv,\yv)$ to
denote the joint distribution of transmission pattern $\xv$ and
feedback information vector $\yv$ under $G_D$ and $G_{Hi}$, $0\leq
i\leq M$. We use $\Pd(\Ac;G_{Hi})$ to denote the probability of event
$\Ac$ under distribution $P_i(\xv,\yv)$.  Note that $\Pd(\Ac;G_{Hi})$
implicitly depends on the underlying direct interference graph $G_D$.

\subsubsection{Construct $G_D$ and $G_{H0}$}
Assume $s\geq 2$. We now construct an underlying direct interference
graph $G_D$, and add directed edges to form $G_{H0}$.  An illustrative
example of a possible $G_D$ is provided in
Fig.~\ref{fig:int_example4}. We partition the node set into $\lceil
n/(2(d+1)+s-1)\rceil $ groups. The first $\lfloor
n/(2(d+1)+s-1)\rfloor$ groups consist of $2d+s-1$ nodes.  The last
group consists of the remaining nodes. In each group, except the last,
we cluster $2(d+1)$ nodes into a pair of cliques, each clique
consisting of $d+1$ nodes.  The remaining $s-1$ nodes are
``independent'' nodes or ``atoms'', disconnected from all other nodes
in the network.  Thus, their activation statuses depend only on their
own queue statuses; they are independent of everything else.

We construct $G_{H0}$ by adding directed edges to $G_D$.  These edges
will be added only between nodes in the same group.  Hidden
interference will thus exist only among nodes within the same group.
It will not exist between groups.  To construct the hidden interference $G_{H0}$ consider each
node in each clique in each group.  Let all $s-1$ independent nodes in
that same group be hidden interferers as well as one (any one) node in
the other clique in the same group.  Thus every node in each clique
has exactly $s$ hidden interferers.  We note that a node in a clique
is allowed to interfere with more than one node in the other clique.
The last group can have an arbitrary edge structure as long as it
satisfies the maximum degree constraints.

Part of the hidden graph structure are the probabilities $p_{ij}$,
defined in~(\ref{eqn:pij}).  Recall that $p_{ij}$ tells us the
likelihood that hidden interferer $i \in \Sc_j$ interferes with the
transmission of node $j$.  We now specify these probabilities for
$G_{H0}$.  For all $i\in \Sc_j$ and $\Sc\subseteq
\Vc\backslash\{i,j\}$, the hidden interferers satisfy
\begin{align}
&\Pd(Y_j=0|X_i=1,X_j=1,\Xv_{\Vc\backslash\{i,j\}}=\ov ;
  {G_{H0}})\nonumber\\ &=\Pd(Y_j\!=\!0|X_i\!=\!1,X_j\!=\!1,\Xv_{\Sc}\!=\!\xv_{\Sc};
  {G_{H0}})=p_{\min}, \label{eqn:pij_equal}
\end{align}
where {$\xv_{\Sc}$} is any transmission pattern feasible under the
direct interference graph $G_D$.  Thus, in contrast to the
inequality~(\ref{eqn:pij_bound}) in the general setting, for this
network the bound holds with {\it equality} for all $i\in \Sc_j$.  The
implication is that the transmission collision probability for an AP
$j$ doesn't increase if there is more than one hidden interferer
transmitting. This assumption holds for every hidden interference
graph $G_{Hi}$ discussed in this section.

\subsubsection{Construct $M$ hidden interference graphs}

We now construct a set of hidden interference graphs $G_{H1},
G_{H2},\ldots, G_{HM}$ as perturbations of $G_{H0}$.  We construct
each graph by removing a single directed edge in $G_{H0}$.  The edge
we remove connects a pair of nodes that are in distinct cliques in a
single group in $G_{H0}$.  To get the graph we do {\em not} remove an
edge between an independent node and a node in a clique.

In each group, there are $2(d+1)$ such edges.  There are thus $2 (d+1)
\lfloor n/(2(d+1)+s-1)\rfloor$ distinct edges in {$G_{H0}$} that can
be removed. If
\begin{align}
d+1\leq c_1n, \quad s-1\leq c_2n,\quad \mbox{and} \quad 2c_1+c_2<1,
\end{align}
where $c_1,c_2$ are positive constants, then there are more than
$2c_1(\frac{1}{2c_1+c_2}-1)n:=M$ such edges. For each of these graphs,
$D_L(D_{H0},D_{Hi})=1$ and $D_L(D_{Hi},D_{Hj})= 2$ where $1 \leq
i,j\leq M$.
\begin{figure}
\centerline{ \includegraphics[width=6cm]{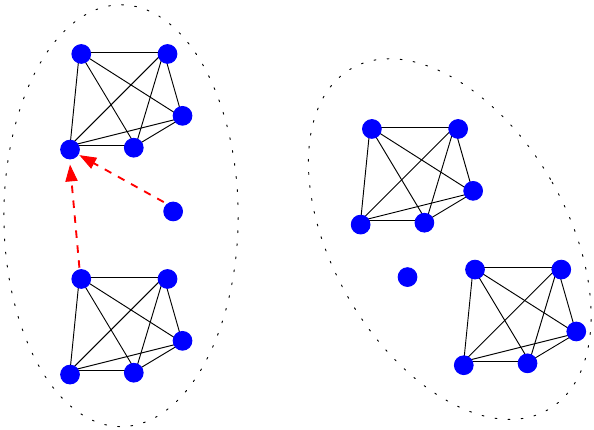}}
\caption{A direct interference graph $G_D$ of $ n/(2(d+1)+s-1) = 2 $
  groups, where each group consists of two fully connected cliques of
  size $d+1$ and $s-1$ detached APs; Dashed arrows indicate the hidden
  interferers of one node; $n=22,d=4,s=2$.}
\label{fig:int_example4}
\vspace*{-0.15in}
\end{figure}

\subsubsection{Characterize $D_{KL}(P_1\|P_0)$}
Without loss of generality, we assume that one directed edge $(i, j)$
{is removed from $G_{H0}$} to form $G_{H1}$, where $i\in\Cc_1$, and
$j\in\Cc_2$.  Since we are {\em removing} an edge from $G_{H0}$ to get
$G_{H1}$, the probability defined in~(\ref{eqn:pij_equal}) is
  inherited. Thus, the only difference between the distributions
under $G_{H1}$ and $G_{H0}$ occurs when $\Xv_{\Cc_1}=\ev_i,
\Xv_{\Cc_2}=\ev_j$, and $\Xv_{\Sc_j\backslash i}=\ov$. Specifically,
\begin{align*}
  \Pd(Y_j=0|\Xv_{\Cc_1}=\ev_i, \Xv_{\Cc_2}=\ev_j,\Xv_{\Sc_j\backslash i}=\ov;G_{H0})&=p_{\min},
\end{align*}
while
\begin{align*}
  \Pd(Y_j=0|\Xv_{\Cc_1}=\ev_i, \Xv_{\Cc_2}=\ev_j,\Xv_{\Sc_j\backslash i}=\ov;G_{H1})&=0.
\end{align*}
Therefore,
\begin{align*}
    D_{KL}&(P_1\|P_0)=\sum_{\xv,\yv}P_1(\xv,\yv)\log \frac{P_1(\xv,\yv)}{P_0(\xv,\yv)}\\
    &=\sum_{\xv,\yv}P_1(\xv)P_1(\yv|\xv)\log \frac{P_1(\yv|\xv)P_1(\xv)}{P_0(\yv|\xv)P_0(\xv)}\\
 &=\Pd(\Ac;G_{H1})\sum_{y_j\in\{0,1\}}P_1(y_j|\Ac)\log \frac{P_1(y_j|\Ac)}{P_0(y_j|\Ac)}\\
    &= -\left(\frac{1-(1-p)^{d+1}}{d+1}\right)^2(1-p)^{s-1}\log (1-p_{\min})
\end{align*}
where $P_1(\xv) = P_0(\xv)$ since $G_{D}$ is held fixed,
$\Ac:=\{\Xv_{\Cc_1} = \ev_i, \Xv_{\Cc_2}=\ev_j,\Xv_{\Sc_j\backslash
  i}=\ov$\}. $\Pd(\Ac;G_{H1})=\Pd(\Xv_{\Cc_1}=\ev_i;G_{H_1})\Pd(
\Xv_{\Cc_2}=\ev_j;G_{H1})\Pd(\Xv_{\Sc_j\backslash i}=\ov;G_{H1})$
because $\Cc_1$, $\Cc_2$ and $\Sc_j\backslash i$ are disconnected in
$G_D$. $\Pd(\Ac;G_{H1})$ is then calculated following a similar
sequence of steps as used to obtain~(\ref{eq.P0dist}).

\subsubsection{Put the pieces together}
In summary, we have $M=2c_1(\frac{1}{2c_1+c_2}-1)n$ and
$D_L(G_{Hi},G_{Hj})\geq 1$ for $0\leq i<j\leq M$.
Theorem~\ref{thm:low3} is proved by an application of
Theorem~\ref{thm:minimax}.


\begin{thebibliography}{10}
\bibitem{yang_isit12}
J.~Yang, S.~Draper, and R.~Nowak, ``Passive learning of the interference graph
  of a wireless network,'' \emph{IEEE International Symposium on Information
  Theory}, pp. 2735--2740, July 2012.

\bibitem{int_map}
D.~Niculescu, ``Interference map for 802.11 networks,'' in \emph{Proceedings of
  the 7th ACM SIGCOMM Conference on Internet Measurement}, 2007, pp. 339--350.

\bibitem{int_probe}
N.~Ahmed and S.~Keshav, ``Smarta: A self-managing architecture for thin access
  points,'' in \emph{Proceedings of the 2006 ACM CoNEXT Conference}, 2006, pp.
  9:1--9:12.

\bibitem{LiuEtAl:13}
B.~N. L.~Liu, Y.~Li and Z.~Pi, ``Heterogeneous cellular networks,'' in
  \emph{Radio Resource and Interference Management for Heterogeneous Networks},
  R.~Q. Hu and Y.~Qian, Eds.\hskip 1em plus 0.5em minus 0.4em\relax Jon Wiley
  \& Sons, 2013, ch.~2.

\bibitem{vivek_thesis}
V.~V. Shrivastava, ``Optimizing enterprise wireless networks through
  centralization,'' Ph.D. dissertation, University of Wisconsin--Madison, 2010.

\bibitem{pie_nsdi}
V.~Shrivastava, S.~Rayanchu, S.~Banerjee, and K.~Papagiannaki, ``{PIE} in the
  sky: Online passive interference estimation for enterprise {WLAN}s,'' in
  \emph{Proceedings of the 8th USENIX Conference on Networked Systems Design
  and Implementation}, 2011, pp. 337--350.

\bibitem{jigsaw1}
Y.-C. Cheng, J.~Bellardo, P.~Benk\"{o}, A.~C. Snoeren, G.~M. Voelker, and
  S.~Savage, ``Jigsaw: Solving the puzzle of enterprise 802.11 analysis,''
  \emph{ACM SIGCOMM}, vol.~36, no.~4, pp. 39--50, Aug. 2006.

\bibitem{jigsaw2}
Y.-C. Cheng, M.~Afanasyev, P.~Verkaik, P.~Benk\"{o}, J.~Chiang, A.~C. Snoeren,
  S.~Savage, and G.~M. Voelker, ``Automating cross-layer diagnosis of
  enterprise wireless networks,'' \emph{ACM SIGCOMM}, vol.~37, no.~4, pp.
  25--36, Aug. 2007.

\bibitem{wit}
R.~Mahajan, M.~Rodrig, D.~Wetherall, and J.~Zahorjan, ``Analyzing the
  {MAC}-level behavior of wireless networks in the wild,'' \emph{ACM SIGCOMM},
  vol.~36, no.~4, pp. 75--86, Aug. 2006.

\bibitem{Castro04}
R.~Castro, M.~Coates, G.~Liang, R.~Nowak, and B.~Yu, ``Network tomography:
  {R}ecent developments,'' \emph{Statistical Science}, vol.~19, pp. 499--517,
  2004.

\bibitem{Duffield:2002}
N.~G. Duffield, J.~Horowitz, F.~L. Presti, and D.~F. Towsley, ``Multicast
  topology inference from measured end-to-end loss,'' \emph{IEEE Transactions
  on Information Theory}, vol.~48, no.~1, pp. 26--45, 2002.

\bibitem{Castro:2004:LBH}
R.~Castro, M.~Coates, and R.~Nowak, ``Likelihood based hierarchical
  clustering,'' \emph{IEEE Transactions on Signal Processiong}, vol.~52, no.~8,
  pp. 2308--2321, Aug. 2004.

\bibitem{SattariFM13}
P.~Sattari, C.~Fragouli, and A.~Markopoulou, ``Active topology inference using
  network coding,'' \emph{Physical Communication}, vol.~6, pp. 142--163, 2013.

\bibitem{AtiaS12}
G.~Atia and V.~Saligrama, ``Boolean compressed sensing and noisy group
  testing,'' \emph{IEEE Transactions on Information Theory}, vol.~58, no.~3,
  pp. 1880--1901, 2012.

\bibitem{Cheraghchi:2012:GGT}
M.~Cheraghchi, A.~Karbasi, S.~Mohajer, and V.~Saligrama, ``Graph-constrained
  group testing,'' \emph{IEEE Transactions on Information Theory}, vol.~58,
  no.~1, pp. 248--262, Jan. 2012.

\bibitem{TosIc:2013:DSF}
T.~To\v{s}i\'{c}, N.~Thomos, and P.~Frossard, ``Distributed sensor failure
  detection in sensor networks,'' \emph{Signal Processing}, vol.~93, no.~2, pp.
  399--410, Feb. 2013.

\bibitem{Lin:2015:OUA}
Y.~Lin, W.~Bao, W.~Yu, and B.~Liang, ``Optimizing user association and spectrum
  allocation in {HetNets}: {A} utility perspective,'' \emph{IEEE Journal on
  Selected Areas in Communications}, vol.~33, no.~6, pp. 1025--1039, June 2015.

\bibitem{Simeone:16}
O.~Simeone, A.~Maeder, M.~Peng, O.~Sahin, and W.~Yu, ``Cloud radio access
  network: {V}irtualizing wireless access for dense heterogeneous systems,''
  \emph{Journal on Communnications and Networks}, vol.~18, no.~2, pp. 135--149,
  April 2016.

\bibitem{ieee802}
\emph{{IEEE} 802.11 standard}, http://www.ieee802.org/11/.

\bibitem{karp}
R.~M. Karp, \emph{Reducibility Among Combinatorial Problems}.\hskip 1em plus
  0.5em minus 0.4em\relax New York: Plenum, 1972.

\bibitem{int_graph_arxiv}
J.~Yang, S.~C. Draper, and R.~D. Nowak, ``Learning the interference graph of a
  wireless network,'' \emph{CoRR}, vol. abs/1208.0562, 2012. [Online].
  Available: http://arxiv.org/abs/1208.0562

\bibitem{pc_dag}
M.~Kalisch and P.~B\"{u}hlmann, ``Estimating high-dimensional directed acyclic
  graphs with the {PC}-algorithm,'' \emph{Journal of Machine Learning
  Research}, vol.~8, pp. 613--636, May 2007.

\bibitem{Tsybakov2008}
A.~Tsybakov, \emph{Introduction to nonparametric estimation}.\hskip 1em plus
  0.5em minus 0.4em\relax Springer, 2008.
  \end{thebibliography}
\end{document}